# *PhyloPythiaS+:* A self-training method for the rapid reconstruction of low-ranking taxonomic bins from metagenomes

**I. Gregor[1,2], J. Dröge[1,2], M. Schirmer[3], C. Quince[3], A. C. McHardy[1,2,4*]**

[1]Max-Planck Research Group for Computational Genomics and Epidemiology, Max-Planck Institute for Informatics, University Campus E1 4, 66123 Saarbrücken, Germany

[2]Department of Algorithmic Bioinformatics, Heinrich-Heine-University Düsseldorf, Universitätsstr. 1, 40225 Düsseldorf, Germany

[3]School of Engineering, University of Glasgow, Glasgow G12 8LT, Scotland

[4]Computational Biology of Infection Research, Helmholtz Center for Infection Research, Inhoffenstraße 7, 38124 Braunschweig, Germany

[*]**Correspondence to:** mchardy@hhu.de

## Abstract

Metagenomics is an approach for characterizing environmental microbial communities *in situ,* it allows their functional and taxonomic characterization and to recover sequences from uncultured taxa. For communities of up to medium diversity, e.g. excluding environments such as soil, this is often achieved by a combination of sequence assembly and binning, where sequences are grouped into 'bins' representing taxa of the underlying microbial community from which they originate. Assignment to low-ranking taxonomic bins is an important challenge for binning methods as is scalability to Gb-sized datasets generated with deep sequencing techniques. One of the best available methods for the recovery of species bins from an individual metagenome sample is the expert-trained *PhyloPythiaS* package, where a human expert decides on the taxa to incorporate in a composition-based taxonomic metagenome classifier and identifies the 'training' sequences using marker genes directly from the sample. Due to the manual effort involved, this approach does not scale to multiple metagenome samples and requires substantial expertise, which researchers who are new to the area may not have. With these challenges in mind, we have developed *PhyloPythiaS+*, a successor to our previously described method *PhyloPythia(S)*. The newly developed + component performs the work previously done by the human expert. *PhyloPythiaS+* also includes a new *k*-mer counting algorithm, which accelerated *k*-mer counting 100-fold and reduced the overall execution time of the software by a factor of three. Our software allows to analyze Gb-sized metagenomes with inexpensive hardware, and to recover species or genera-level bins with low error rates in a fully automated fashion.

**Availability**: A distribution of *PPS+* in a virtual machine is available for installation under Windows, Unix systems or OS X. The software is available on: https://github.com/algbioi/ppsp/wiki

# 1   Extended abstract

Metagenomics is an approach for characterizing environmental microbial communities *in situ,* it allows their functional and taxonomic characterization and to recover sequences from uncultured taxa. A major aim is to reconstruct (partial) genomes for individual community members from metagenomes. For communities of up to medium diversity (e.g. excluding environments such as soil), this is often achieved by a combination of sequence assembly and binning, where sequences are grouped into 'bins' representing taxa of the underlying microbial community from which they originate. If sequences can only be binned to higher-ranking taxa than strain or species, these bins offer less detailed insights into the underlying microbial community. Therefore, assignment to low-ranking taxonomic bins is an important challenge for binning methods as is scalability to Gb-sized datasets generated with deep sequencing techniques. Due to the importance of a match of the training data to the test data set in machine learning for achieving high classification accuracy, one of the best available methods for the recovery of species bins from an individual metagenome sample[1,2] is the expert-trained *PhyloPythiaS* package, where a human expert identifies the 'training' sequences directly from the sample using marker genes and contig coverage information and based on data availability decides on the taxa to incorporate into the composition-based taxonomic model. The sequences of a metagenome sample are consequently assigned to these or higher ranking taxa by *PhyloPythiaS.* Because of the manual effort involved, this approach does not scale to multiple metagenome samples and requires substantial expertise, which researchers who are new to the area may not have. Other methods for draft genome reconstruction use multiple related metagenome samples as input[3,4] or are not distributed as a software package[5].

With these challenges in mind, we have developed *PhyloPythiaS+*, a successor to our previously described method *PhyloPythia(S)*[2,6]. The newly developed + component performs the work of the human expert (Section 3). It screens the metagenome sample for sequences carrying copies of one of 34 taxonomically informative marker genes[7] (Section 4.3). Identified marker genes are taxonomically classified using an extensive reference gene collection. The + component then decides which taxa to incorporate into the composition-based taxonomic model based on the amount of available sequence data identified from the metagenome sample, genome and draft genome reference sequence collections (Figure 4).

We evaluated *PhyloPythiaS+* on metagenome datasets of assembled simulated reads with Illumina GAII error profiles generated from a log-normal or uniform abundance distribution over 47 strains, and two real metagenome datasets from human gut and cow rumen samples (Tables 10–13, Sections 4 and 5). *PhyloPythiaS+* had substantially higher overall precision and recall than the generic *PhyloPythiaS* model, because of the better match of the composition-based taxonomic model to the sequenced microbial community (Figs 1 and 5–8, Section 4.9). It performed similarly well to an expert-trained *PhyloPythia* model without requiring manual effort (Figure 2, Table 14). Comparisons to sequence-similarity-based methods such as the popular MEtaGenome ANalyser (MEGAN, version 4)[8] and our own *taxator-tk*[9] software showed a substantial increase in correct assignments to low taxonomic ranks for *PhyloPythiaS+,* while maintaining acceptably low error rates (Figs 2 and 5–9). The largest improvement in comparison to the other methods was observed for taxa from deep-branching lineages, such as from genera or families without sequenced genomes but with marker gene data for the strain or species available (Fig. 5–8, Table 9: Test Scenarios 2–4). This is currently the case for 39,201 species represented in our 16S reference gene collection.

*PhyloPythiaS+* includes a new *k*-mer counting algorithm based on the Rabin Karp string matching algorithm (Section 3.3). The algorithm accelerated *k*-mer counting 100-fold and reduced the overall execution time of the software by a factor of three in comparison to the original *PhyloPythiaS* release (Figure 3). We found that 500 and 360 Mb/hour could be assigned by *PhyloPythiaS+* on a single CPU core of a standard compute server and a laptop, respectively. Our software thus allows to analyze Gb-sized metagenomes with inexpensive hardware, and to recover species or genera-level bins with low error rates in a fully automated fashion. *PhyloPythiaS+* is distributed in a virtual machine and is easy to install for all common operating systems (Section 6).

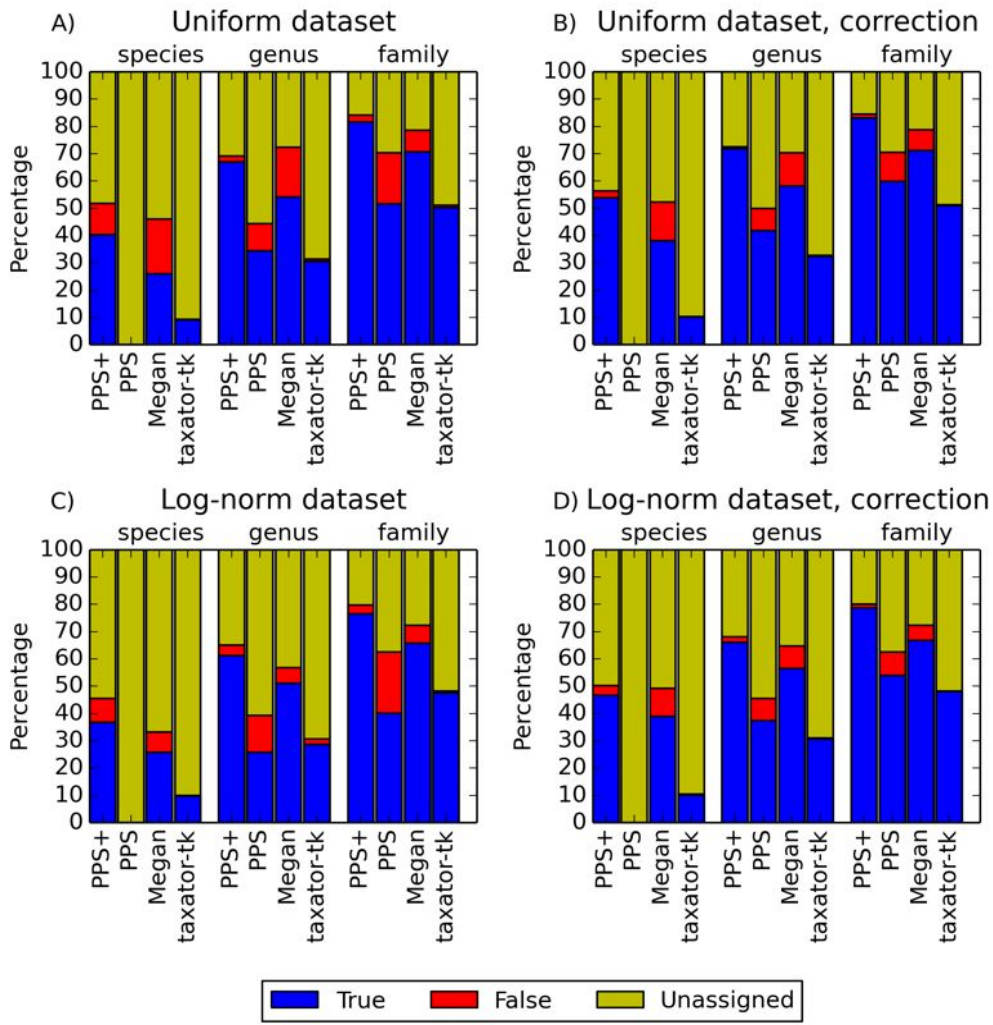

**Figure 1: Performance comparisons with simulated datasets.**

Panels A and C show the fraction of correct, false and unassigned bp for simulated datasets with uniform and log-normally distributed species abundance for *PhyloPythiaS+*, the generic *PhyloPythiaS* model, *MEGAN4* and *taxator-tk* for assignments at the species, genus and family ranks. Results were averaged over seven 'scenarios', where sequences of the same strain, species or genus from the simulated metagenomes were removed from the genome, draft genome and marker gene reference sequence collections (Figs 5 and 7). Panels B and D show the portion of consistently, inconsistently and unbinned bp without consideration of the taxonomic identifiers (Figs 6 and 8, Section 4.9).

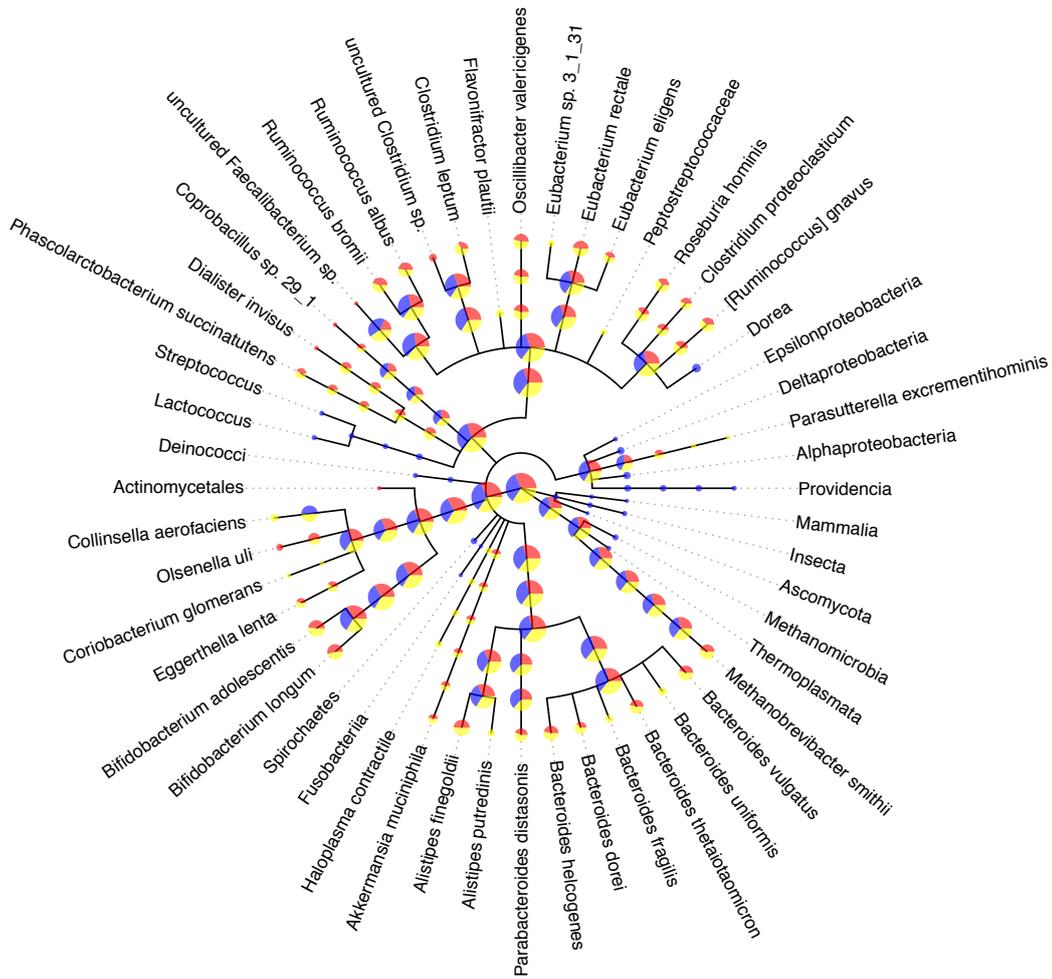

**Figure 2: Comparison to expert binning based on marker genes.**

The amount of assigned bp by *PhyloPythia* (blue), *PhyloPythiaS+* (yellow) and taxonomically informative marker genes directly (red) to each taxon are indicated by the pie chart sizes on a log-scale for two human gut metagenome samples[2,10]. *PhyloPythiaS+* automatically determined the taxa to model from the samples. For the expert-trained *PhyloPythia*, the taxa to model were specified by an expert, and were included in the model if they were covered by sufficient reference sequence data retrieved separately from the sample and from sequenced human gut isolates. *PhyloPythiaS+* assigned sequences to low-ranking taxa down to the species level, in agreement with the marker gene assignments, while *PhyloPythia* often assigned these sequences to the parental taxa. *PhyloPythia* also included eukaryotes in the model.

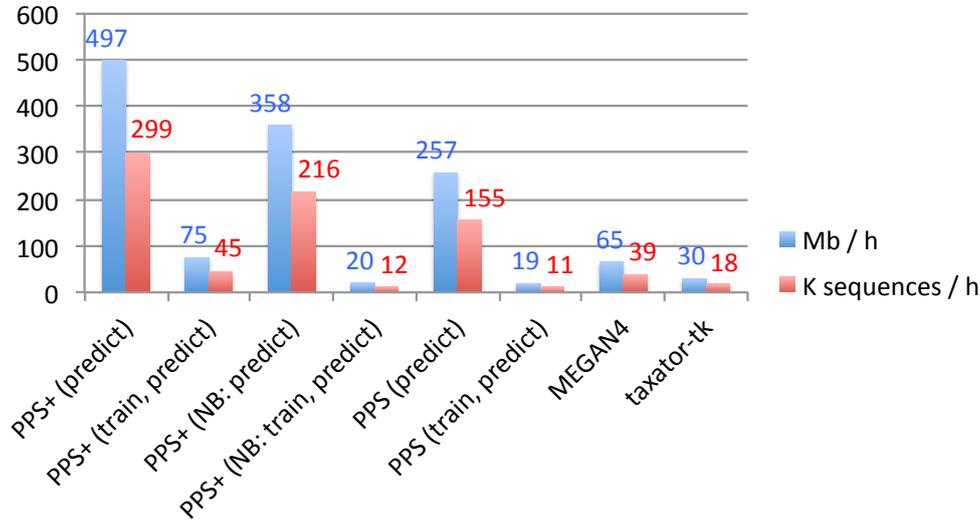

**Figure 3: Empirical comparison of execution times.**

The throughput was measured in Mb and the number of sequences classified within 1 hour with one execution thread, using all assembled contigs of the human gut metagenome dataset on a server computer with an AMD Opteron 6386 SE 2.8 GHz processor and 512 GB of RAM. Default settings were used for all methods (Sections 4.5–4.8). Both *MEGAN4* and *taxator-tk* were run using *BLAST*. For *MEGAN4,* only the runtime of *BLAST* was considered, as the runtime of the subsequent algorithm was negligible. For *PhyloPythiaS* and *PhyloPythiaS+*, the throughput was calculated for the prediction phase and both phases (training and prediction). The former is relevant when using previously generated models for the classification of multiple samples. The execution time shown for *PhyloPythiaS* is approximately three times better than that for the original release, as we incorporated the new *k*-mer counting algorithm. *PhyloPythiaS+* was the only method that could also be executed on a standard laptop with an Intel i5 M520 2.4 GHz processor and 4 GB of RAM.

## *Table of Contents*

## *Tables*

## *Figures*

# 2 Introduction

Metagenomics is the functional or sequence-based analysis of microbial DNA isolated directly from a microbial community of interest[11,12]. As the cultivation conditions for most microorganisms are unknown or too complex to reproduce in the laboratory[13], random shotgun and amplicon-sequencing based metagenome studies have led to substantial advances in our understanding of the structure and functions of microbial communities within the last decade (see, for instance[1,10,14-18]). With the simultaneous rapid advances of the sequencing technologies in terms of increasing throughput and decreasing costs[19,20], computational efficiency has become an essential requirement for metagenome analysis methods.

The taxonomic classification or 'binning' of random shotgun metagenome samples complements sequence assembly methods in returning sets of sequence fragments with the same taxonomic identifier, which represent draft genomes or pan-genomes of microbial community members. The subsequent analysis of these bins allows to characterize the functional and metabolic potential for individual member species instead of performing a 'bag of genes' analysis for the entire sample. In a recent collaboration with Mark Morrison's group, a functional and metabolic analysis of a draft genome recovered by taxonomic binning from the metagenome of a microbial community from the gut of the Australian Tammar Wallaby led to the isolation and subsequent sequencing of a new and previously uncultivated bacterium[1].

While taxonomic binning generates an extensive assignment of a shotgun metagenome sample for subsequent functional genomic analysis of the recovered bins, in taxonomic profiling a small subset of the genes of a shotgun sample are classified to determine a taxonomic profile for the underlying community. Profiling tools such as *AMPHORA*[21], *AMPHORA2*[7], *MLTreeMap*[22] and mOTU[23] perform taxonomic assignment based on up to 140 taxonomically informative marker genes, while *MetaPhlAn*[24] uses 400,141 clade-specific marker genes for this purpose.

Taxonomic binning methods use sequence homology, sequence composition in terms of short oligomer usage (known as the 'genome signature') and similarities in sequence coverage or gene counts, individually or in combination for taxonomic assignment[25]. Homology-based methods analyze local sequence similarities of a sample sequence to reference sequences of known taxonomic origin for taxonomic assignment. Initially, sequence similarity searches against a reference sequence collection are performed to find homologs for the sample sequences. Subsequently, the algorithms implemented in *taxator-tk*[9], *MEGAN4*[8], *CARMA3*[26], or *SOrt-ITEMS*[27] derive a taxonomic assignment from the taxonomic identifiers of the closest homologs under consideration of their location in a reference taxonomy. Homology-based methods are more accurate than composition-based methods in assigning over large taxonomic distances and for short fragments of few 100 bp in length. However, they only allow low-level taxonomic assignment, if genome sequences from closely related organisms to the members of the sequenced microbial community are available. This is the case for approximately 40% of the members of the human microbiota through efforts of the Human Microbiome Project[28], but for few species from less well studied environments. Furthermore, the approach can be computationally very demanding, as the runtime increases proportional to the size of the reference sequence collection multiplied by the size of the metagenome sample.

Composition-based methods assign metagenome sequences based on their k-mer signature, which is derived from the counts of short oligomers (k-mers) for a sequence[29,30]. Methods such as *PhyloPythia*[6], *PhyloPythiaS (PPS)*[2], *PhymmBL*[31] or the Naïve Bayes classifier *(NBC)*[32] use this property for taxonomic sequence assignment. Contrary to sequence homology, composition-based signatures are global genomic properties, which can be estimated from any sufficiently sized sequence sample for a taxon; e.g. for *PhyloPythia(S)*, 100 kb of reference sequences for a taxon are sufficient for accurate assignment in particular for low ranking taxa, and no complete genome sequences are required. This is a very important property for metagenome analysis, as thus no complete genomes of the same or related species are needed as reference information, which oftentimes are not available. As they furthermore do not rely on similarity searches against large sequence collections, composition-based methods tend to be fast, with their runtimes oftentimes increasing linearly with the size of the metagenome sample. As the assignment accuracy of composition-based techniques is reduced for assignments of fragments of less than 1 kb, for *PhyloPythia* and its successors the assignment of longer fragments is recommended, making it suitable for assembled data sets and reads generated with PacBio[33] sequencing technology.

As a third category, coverage-based approaches make use of the observation that contigs with similar coverage in assembly are likely to originate from the same population of a community[4]. Optionally in combination with composition-based analysis, this can be used for the recovery of draft genomes from deeply branching lineages for which no related genome sequences are available[5]. Additionally, taxonomic bins can be reconstructed based on variation in gene counts or contig coverage, if multiple samples are available[3,34].

Taxonomic classification methods should be able to efficiently and accurately assign Gb-sized random shotgun metagenome samples, as currently a single run of Illumina HiSeq 2500 produces up to 600 Gb of sequence data in 11 days. We have developed a fully automated taxonomic binning method that can process Gb-sized samples on hardware available to most users. *PhyloPythiaS+* (*PPS+*) extends our previously described *PhyloPythiaS* (*PPS*) software. We are aware of only a few taxonomic binning methods, namely *PPS*, *taxator-tk*, and *MEGAN4,* that can process gigabytes of metagenome data using multiple cores per day on standard hardware, and in comparison to these, *PPS+* delivers an additional substantial decrease in processing time. *PPS* uses structured support vector machines for the composition-based taxonomic classification of metagenome samples[35], which improves assignment consistency for higher taxonomic ranks. The recommended use of *PPS* delivering the best results is that a human expert first decides which taxa to include in the model, based on available sequences for a taxon and knowledge of taxa present in the community. Additional information, such as amplicon sequence data sets for marker genes can be considered for identification of the most relevant taxa to be modeled. Training sequences may be derived directly from the sample based on marker gene taxonomic profiling, from similarities in contig coverage information, and in addition from public sequence collections. Sample-derived *PPS* models are then trained for the selected taxa and are subsequently used for the taxonomic classification of the metagenome sample. Without expert involvement, a generic model including all taxa with multiple genome sequences can be used for taxonomic classification. However, the latter results in substantially decreased assignment accuracy for samples of microbial communities with few members from extensively sequenced clades. We therefore developed a fully automated training procedure and implemented this in *PPS+* to allow accurate metagenome assignment to low ranking taxa without manual intervention or consideration of additional information. *PPS+* is self-training and infers all required information from the metagenome sample itself by taxonomic profiling of marker genes and scanning of reference sequence collections. If wanted, it is still possible to manually adjust the input to *PPS*. An initial version of *PPS+* has been already employed by Pope *et al.*[36].

In the following, we describe *PhyloPythiaS+* (*PPS+*) and its evaluation in detail. We begin by shortly introducing the *PhyloPythiaS* (*PPS)* taxonomic classifier (Section 3.1) and then describe its extension by the + component, which allows to determine suitable training data for low-ranking taxonomic bins from deep-branching lineages based on marker gene analyses (Section 3.2). This is followed by a description of our new and accelerated *k*-mer counting algorithm (Section 3.3), that is used to generate the feature vectors of oligomer frequencies from DNA sequences for training of the structured Support Vector Machines used in *PPS+*. We then describe the benchmark settings, i.e. how we evaluated the performance of *PPS+* (Section 4). We begin with outlining the design of the simulated metagenome datasets (Section 4.1) and specifying the real datasets that we used in the evaluation (Section 4.2). We describe the reference sequence collections used for benchmarking all tools (Section 4.3), the corresponding hardware (Section 4.4) and the configurations of all tools (Sections 4.5–4.8). We end this section with the definitions of the evaluation measures, i.e. the micro-averaged precision and recall, which we used to evaluate the benchmark experiments with the simulated metagenome datasets (Section 4.9) and the scaffold-contig consistency measures employed to evaluate the benchmark experiments with the real metagenome datasets (Section 4.10). The results section discusses the results obtained for all evaluated tools with the simulated datasets (Sections 5.1 and 5.2), the real datasets (Section 5.3) and an empirical throughput comparison for all tools (Section 5.4). The conclusions section (Section 6) summarizes our main findings.

# 3   Methods

*PPS+* proceeds in two phases: In the first phase, the newly developed + component identifies sample-derived training sequences and the taxa to be modeled by searching for copies of 34 ubiquitous taxonomic marker genes in the metagenome sample (Figure 4). The second phase is the composition-based taxonomic assignment of the sample using *PPS*. The marker gene analysis results in taxonomic assignments for a small fraction of the metagenome sample sequences. Based on the taxa abundance profile derived from the marker gene assignments and the sequences available in the reference sequence collections, our method heuristically determines which taxa will be modeled and which are the sample-derived data that will be used for training *PPS*. The *PPS* models are inferred from the training data and are subsequently used for taxonomic classification of the entire metagenome sample. In a post-processing step, multiple evaluation measures are computed to assess the quality of the generated taxonomic binning. *PPS* models can optionally be reused to classify further metagenome samples, such as additional samples from the same community.

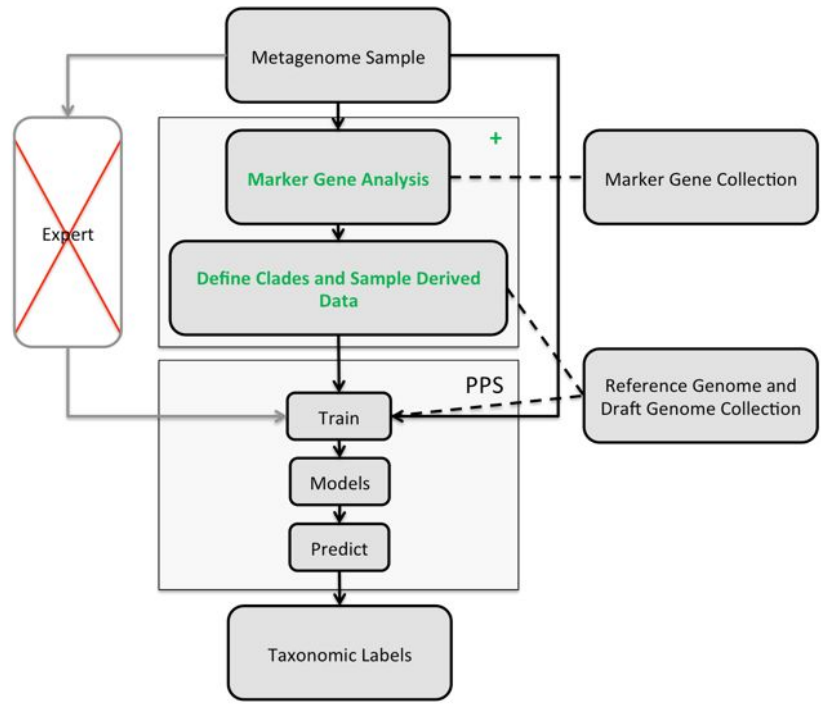

**Figure 4: *PPS*+ workflow.**

The standard use of *PPS* is that a human expert decides the taxa to model and which sample-derived data to use for *PPS* training, oftentimes including marker-gene carrying contigs from the metagenome sample. In *PPS*+, the + component performs this task using marker genes that it identifies from the sample and (draft) genome sequence collections. *PPS* models for the selected taxa are then trained using these sequence data and subsequently applied to taxonomically assign the metagenome sequence sample.

## 3.1   PhyloPythiaS

Assignment with *PPS* has two phases: In the training phase, structured output Support Vector Machines (SVMs) for the specified part of the NCBI taxonomy, defined by the taxa being modeled, are trained using the sample-derived training sequences and additional data for these taxa from a customized reference collection of sequenced genomes and draft genomes (Section 4.3). The output is an ensemble of structured output models for the input taxonomy, which can be used for the taxonomic classification of a metagenome sequence sample.

For the training phase, the list of leaf node taxa and sample-derived data generated with the + component are used as the input for *PPS*. The list of clades restricts the taxonomic output space that is modeled, i.e. a sequence from a metagenome sample will be assigned to a leaf node taxon or a corresponding higher-ranking taxon.

In the prediction phase, the *PPS* model ensemble identifies the taxon which best matches a query sequence in terms of its k-mer profile and assigns the respective taxonomic identifier to the query sequence. This can be to any internal or leaf node taxon in the learned taxonomy. By default, sequences of 1 kb or more are classified (*PPS+* configuration parameter: *minSeqLen=1000*).

## 3.2   The + component of PhyloPythiaS+

The input for the analysis by the + component of *PhyloPythiaS+* is the metagenome sequence sample. This step returns a list of clades and sample-derived data that are subsequently used to train *PPS* models. The analysis of the + component consists of the following steps:

1) *Marker gene identification*: DNA sequences from the metagenome are translated in all six reading frames (i.e. also considering reverse complement sequences) to protein sequences. In both the translated and untranslated sequences, regions with similarity to the DNA or protein Hidden Markov model (HMM) profiles of 34 taxonomically informative marker genes are identified. The marker gene sequences from these regions are used for further analysis and translated back into DNA sequences in case they correspond to protein sequences. The search command of the *HMMER 3*[37] package was used to identify sequence regions with similarity to the HMM profiles for the following taxonomic marker genes:  16S, 23S, 5S, *dnaG, infC, pgk, rpoB, tsf,* *frr*, *nusA*, *pyrG*, *rpmA*, *smpB*, *rpsC*, *rpsI*, *rpsK*, *rpsS*, *rpsB*, *rpsE*, *rpsJ*, *rpsM*, *rplA*, *rplB*, *rplC*, *rplD*, *rplE*, *rplF*, *rplK*, *rplL*, *rplM*, *rplN*, *rplP*, *rplS* and *rplT*. The *e*-value cut-off for *hmmsearch* was set to 0.01.

2) *Taxonomic marker gene assignment*: The marker gene sequences were assigned a taxonomic identifier using the reimplementation of the composition-based Naïve Bayes taxonomic classifier[38] implemented in *MOTHUR*[39]. Parameters: The number of iterations to compute the bootstrap confidence score was set to 300. The corresponding confidence score cut-off was set to 80. For the 16S marker gene analysis, pieces of code from Huang *et al.*[40] were used.

3) *Taxonomic sequence assignment*: If a sequence fragment contains multiple marker genes, multiple taxonomic identifiers are identified in Step 2. Then the highest bootstrap confidence score (*hcs*) returned by the Naïve Bayes classifier (NBC) for one of the markers on the fragment is identified. We use all marker gene assignments with confidence scores larger than *hcs \* (1 - candidatePlTopPercentThreshold)*. The default setting for the configuration parameter *candidatePlTopPercentThreshold* is 0.1. From the set of taxonomic identifiers, the lowest taxon *t* is identified for which all other assignments are either to the same taxon *t* or defined at higher-ranking parental taxa of *t*. Taxon *t* is consequently used for the overall fragment assignment. The confidence score for the fragment is set to the smallest confidence score for the set of retained marker gene assignments for the following steps.

4) *[Optional: Taxonomic scaffold assignment]*: Scaffolding information can be used to obtain more training data for the relevant taxa. In the first step, the taxonomic identifiers of all assigned contigs for a scaffold are corrected as follows: Let us consider that *n* taxonomically assigned contigs of a scaffold are placed along a common path from the root *r* down to a low-ranking clade *lc* in the reference taxonomy. The unassigned contigs of a scaffold are not among these *n* contigs. To obtain a consistent assignment for all the contigs of a scaffold and to correct for 'outlier' contig assignments to low ranking taxa, contigs are reassigned according to the following: All *n* assigned contigs of the respective scaffold are reassigned to the lowest taxon *c*, which lies on the path from *r* to *lc*, where *c* is chosen such that at least *(agThreshold \* n)* of the contigs are assigned on the path from *c* to *lc*. In the second step, unassigned contigs are assigned to the same taxon *c*, if a sufficient number of contigs have already been assigned. Let us denote the sum of all contig lengths for a scaffold as *l* and the sum of all assigned contig lengths of the respective scaffold as *al*. If $al/l \geq assignedPartThreshold,$ then the unassigned contigs of a scaffold are also assigned to clade *c* (see the configuration parameters: *placeContigsFromTheSameScaffold=True*, *agThreshold=0.3*, *assignedPartThreshold=0.5*).

5) *Assignment path truncation*: Contigs assigned to a lower-ranking taxon than at the species or alternatively specified lowest rank (configuration parameter: *rankIdCut=6*; 6 assignments to the species rank) are reassigned to the parental taxon at this lowest rank.

6) *Taxa selection for model specification*: Any taxon for which at least 100 kb of sample-derived data have been identified can be modeled. Furthermore, species can be modeled if at least 300 kb of reference sequences are available in the reference sequence database, and higher-ranking taxa can be modeled if data for at least three distinct species with this requirement (>300 kb per species) are available. Contigs assigned to taxa for which there are fewer data are subsequently assigned to higher taxonomic ranks for which sufficient data are available to allow their use as sample-derived training data for the taxa that will be included in the model (configuration parameters: *minGenomesWgs=3 or 1*, *minBpPerSpecies=300,000*, *minBpToModel=100,000*).

7) *Abundant taxa selection*: To reduce the number of taxa to the most relevant ones, the least abundant taxon is removed iteratively. This is defined as the taxon to which the minimum number of bp is assigned. Sequences assigned to this taxon are reassigned to the closest defined taxon at a parental rank. The algorithm ends when the number of leaf taxa is less than or equal to the maximum number of taxa to be modeled (configuration parameter: *maxLeafClades=50*; this can also be set to larger values of realistically up to 800).

8) *Balancing training data*: The part of the taxonomy that will be modeled with *PPS* is defined by the taxa identified in the previous step. It has leaf nodes at different ranks above the specified rank cut-off, and internal nodes. Only leaf node taxa and sample-derived training data assigned to leaf node taxa in the preceding steps are specified as input for *PPS* training. To balance the training data across clades, a maximum of 400 kb of sample-derived training data are selected for each leaf node taxon (configuration parameter: *maxSSDfileSize=400,000*). For this selection, contigs are used in order of decreasing confidence values and then in order of decreasing length. The balancing of training data can be switched off by setting the configuration parameter *maxSSDfileSize* to a large number.

## 3.3 The k-mer counting algorithm

### 3.3.1 *Algorithm description*

To accelerate the *k*-mer enumeration in *PPS+*, we replaced the suffix trees used in *PhyloPythiaS*[2] with a custom algorithm based on the Rabin Karp string-matching algorithm[41]. The algorithm is highly optimized to count short DNA sub-strings; moreover it is very fast as it does not need any large helper data structure (similar to suffix trees), explores the locality of reference, uses very fast bit shift operations and is efficiently implemented in C. It enumerated k-mers up to 100 times faster than when using suffix trees, which substantially accelerates taxonomic assignment with *PPS+*.

Let us assume that we are given an array *a*, which represents a DNA sequence of length *n* where all letters are encoded as numbers *0, 1, 2, 3* (where *A≈0, T≈1, G≈2, C≈3*) and let $a_0$, …, $a_{n-1}$ denote the respective entries. We would like to count the occurrences of all *k*-mers of length *k* and store the counts in an array *c* of length $4^k$, which is initialized by zeros. Each *k*-mer maps to a unique index in the array *c*. The index of the first *k*-mer in our sequence is calculated according to

$$index_0 = a_0 * 4^{k-1} + a_1 * 4^{k-2} + \cdots + a_{k-2} * 4^1 + a_{k-1}(* 4^0)$$

The index of the *(i+1)*th *k*-mer of the sequence is computed from the (*i*)[th] index as:

$$index_{i+1} = (index_i - a_i * 4^{k-1}) * 4 + a_{i+k}(* 4^0)$$

When an index is identified, the corresponding $k$-mer count at this index position in array $c$ is incremented by one.

For instance, the DNA sequence *ATGCATG* is encoded in array $a$ as [0, 1, 2, 3, 0, 1, 2]. For $k$=2, we would add two counts for the $k$-mer *AT* in array $c$ at the index position 0*4 + 1 = 1, two counts for *TG* at the index position *1\*4 + 2 = 6,* one count for *GC* at the index position 2*4 + 3 = 11 and one count for *CA* at index position 3*4 + 0 = 12. The multiplication operation $X * 4^m$ can be computed using the bit shift operation $X << 2*m$ (for $m$=0, 1, 2... and $X$=0, 1, 2, …), which is usually much faster than multiplication.

### 3.3.2   Counting k-mers of different lengths at once

If $index_i$ is the index of the $i^{\text{th}}$ $k$-mer of length $k$, the index of the $i^{\text{th}}$ $(k-j)$-mer (of length $k-j$) can be simultaneously computed using the bit shift operation as *$index_i >> (2*j)$* (for $j \in [1 \mathinner{.\,.} k-1]$) and the corresponding counter at the computed index of a respective counter array of length $4^{(k-j)}$ is incremented. The end of a DNA sequence can be handled by adding several non-DNA characters to its end.

### 3.3.3   Non-DNA character handling

In parallel to array $a$, we implicitly maintain array $b$, where $b_i$ is 0 if $a_i$ is a DNA character and $b_i$ is 1 if $a_i$ is a non-DNA character. If $b_i$ is 1 then $a_i$ is 0 and this position is not considered in the $k$-mer computation, so any $k$-mer containing a non-DNA character is not counted and is ignored. Thus, in parallel to the variable $index_i$ for $a$, we also maintain $indexB_i$ for $b$, which is a bitmap. If $indexB_i$ is not 0 then $index_i$ represents a $k$-mer that contains at least one non-DNA character and thus this $k$-mer is not counted.

### 3.3.4   Complexity

The main advantage of our method is that we do not use additional helper data structures similar to suffix trees, since we work directly with arrays that represent DNA sequences. The only larger data structure that is necessary is a one-dimensional array that contains the counts of individual $k$-mers. The algorithm also processes one sequence at a time and thus there is no need to store all the sequences in the main memory, which makes the algorithm memory-efficient. To compute the next index from a previous index, we need to perform only two bit shift operations, one addition, one subtraction and one read operation (of the entry $a_{i+k}$). This ensures complexity $O(n)$, where $n$ is the length of the DNA sequence that is being considered.

### 3.3.5 Evaluation

To evaluate our $k$-mer counting method, we compared it with the state-of-the-art method *Jellyfish*[42]. As a test dataset, concatenated contigs from Turnbaugh *et al.*[10] were used, which amounted to an overall sequence length of 251 Mb. All tests were run in one thread on a computer with an Intel i5 2557M processor (Section 4.4, HW Configuration 5) and SSD storage to minimize the overhead of the I/O operations. In all tests, reverse complement $k$-mers were also identified. The results for both methods are comparable for $k$-mers of lengths up to 12 (Table 1). For longer $k$-mers, *Jellyfish* is substantially faster due to the use of a custom hash table, whereas our algorithm stores the $k$-mers in an ordinary one-dimensional array. The main advantage of our algorithm is that it can calculate the counts of $k$-mers of different lengths at once, which *Jellyfish* cannot. In the setting used in *PPS+*, our algorithm is *~2.5* times faster than *Jellyfish* for computation of $k$-mer length of 4–6. In general, our algorithm can compute $k$-mers of lengths up to *15*, while *Jellyfish* can compute $k$-mers of lengths up to 31.

**Table 1: Runtime comparison of *Jellyfish* and the new k-mer counting algorithm implemented in *PPS+*.**

| *k*- mer lengths | *PPS+* | *Jellyfish* |
|---|---|---|
| 4 | 10.1 s | 10.3 s |
| 5 | 10.1 s | 10.3 s |
| 6 | 10.1 s | 10.4 s |
| 4,5,6 | 11.9 s | N/A |
| 7 | 10.1 s | 10.4 s |
| 8 | 10.1 s | 10.4 s |
| 9 | 10.8 s | 10.8 s |
| 10 | 14.5 s | 14.3 s |
| 11 | 16.7 s | 18.1 s |
| 12 | 21.8 s | 20.1 s |
| 13 | 55.0 s | 23.2 s |

# 4 Benchmark settings

## 4.1 Simulated datasets details and generation

Our simulated mock community comprised 47 strains from 45 different species (37 different genera) defined at all major taxonomic ranks, i.e. at superkingdom, phylum, class, order, family, genus and species rank (Table 2). Two simulated datasets were generated with different abundance profiles, one with a uniform distribution and one with a log-normal distribution ($\mu=1$, $\sigma=2$) (Table 3).

**Table 2: List of strains used to generate simulated datasets.**

| Strain name | Accession number |
| --- | --- |
| *Acidobacterium capsulatum* ATCC 51196 | CP001472.1 |
| *Akkermansia muciniphila* ATCC BAA-835 | CP001071.1 |
| *Archaeoglobus fulgidus* DSM 4304 | AE000782.1 |
| *Bacteroides thetaiotaomicron* VPI-5482 | AE015928.1 |
| *Bacteroides vulgatus* ATCC 8482 | CP000139.1 |
| *Bordetella bronchiseptica* RB50 | BX470250.1 |
| *Caldicellulosiruptor bescii* DSM 6725 | CP001393.1 |
| *Caldicellulosiruptor saccharolyticus* DSM 8903 | CP000679.1 |
| *Chlorobium limicola* DSM 245 | CP001097.1 |
| *Chlorobium phaeobacteroides* DSM 266 | CP000492.1 |
| *Chlorobium phaeovibrioides* DSM 265 | CP000607.1 |
| *Chlorobium tepidum* TLS | AE006470.1 |
| *Chloroflexus aurantiacus* J-10-fl | CP000909.1 |
| *Clostridium thermocellum* ATCC 27405 | CP000568.1 |
| *Deinococcus radiodurans* R1 | AE001825.1<br>AE000513.1 |
| *Dickeya dadantii* 3937 | CP002038.1 |
| *Dictyoglomus turgidum* DSM 6724 | CP001251.1 |
| *Enterococcus faecalis* V583 | AE016830.1 |
| *Fusobacterium nucleatum* subsp. *nucleatum* ATCC 25586 | AE009951.2 |
| *Gemmatimonas aurantiaca* T-27 | AP009153.1 |
| *Herpetosiphon aurantiacus* DSM 785 | CP000875.1 |
| *Hydrogenobaculum* sp. Y04AAS1 | CP001130.1 |
| *Ignicoccus hospitalis* KIN4/I | CP000816.1 |
| *Methanocaldococcus jannaschii* DSM 2661 | L77117.1 |
| *Methanococcus maripaludis* C5 | CP000609.1 |
| *Methanococcus maripaludis* S2 | BX950229.1 |
| *Nitrosomonas europaea* ATCC 19718 | AL954747.1 |
| *Pelodictyon phaeoclathratiforme* BU-1 | CP001110.1 |
| *Persephonella marina* EX-H1 | CP001230.1 |
| *Porphyromonas gingivalis* ATCC 33277 | AP009380.1 |
| *Pyrobaculum aerophilum* str. IM2 | AE009441.1 |
| *Pyrobaculum calidifontis* JCM 11548 | CP000561.1 |
| *Rhodopirellula baltica* SH 1 | BX119912.1 |
| *Ruegeria pomeroyi* DSS-3 | CP000031.1 |
| *Salinispora arenicola* CNS-205 | CP000850.1 |
| *Salinispora tropica* CNB-440 | CP000667.1 |
| *Shewanella baltica* OS185 | CP000753.1 |
| *Shewanella baltica* OS223 | CP001252.1 |
| *Sulfolobus tokodaii* str. 7 | BA000023.2 |
| *Sulfurihydrogenibium* sp. YO3AOP1 | CP001080.1 |
| *Thermoanaerobacter pseudethanolicus* ATCC 33223 | CP000924.1 |
| *Thermotoga neapolitana* DSM 4359 | CP000916.1 |
| *Thermotoga petrophila* RKU-1 | CP000702.1 |
| *Thermotoga* sp. RQ2 | CP000969.1 |
| *Thermus thermophilus* HB8 | AP008226.1 |
| *Treponema denticola* ATCC 35405 | AE017226.1 |
| *Zymomonas mobilis* subsp. *mobilis* ZM4 | AE008692.2 |

**Table 3: Properties of the simulated datasets.**

| Distribution | Contigs | Mb |
|---|---:|---:|
| Uniform | 14,393 | 137 |
| Log-normal | 13,284 | 66 |

A custom read simulator was used which utilizes position- and nucleotide-specific substitution patterns derived from experimental datasets. This allowed us to generate reads with more realistic error profiles than we would with read simulators such as *pIRS*[43], *ART*[44] or *MetaSim*[45]. Furthermore, we could thus specify and test different species abundance distributions for the microbial community and generate very large datasets due to the parallelization of the simulation program. We did not use the simulated datasets from Mavromatis *et al.*[46], as these are substantially smaller than the current metagenome datasets.

Both simulated datasets were generated based on Illumina GAII error profiles where the standard library preparation method was used. The insert size distribution was also based on the experimental dataset. For each dataset, 15 million paired-end reads of 90 bp were generated with an average insert size of 291 bp. The first 10 bp of the 100 bp reads in the experimental dataset were trimmed because of fluctuations in the nucleotide distributions at the starting positions, which indicated partial remains of the barcode sequence. The read simulator produces output in FASTA format, which was converted into a pseudo-FASTQ format for the downstream analysis with uniformly high quality scores. The reads were then assembled with *Metassembler*[47] using *Velvet*[48], run with different *k*-mer sizes ranging between 19 and 75, and were subsequently merged with *Minimus2*[49]. This assembly procedure resulted in a larger assembled dataset than assembly with *SOAPdenovo2*[50], *Metavelvet*[51] or *Newbler*[52]. Contig sequences longer than 1000 bp were considered further. The contigs were subsequently mapped with *BLAST*[53] onto the reference genomes to recover their taxonomic identifiers.

## 4.2   Real datasets

For the evaluation using real metagenome samples from actual microbial communities, we used two metagenome samples from the guts of obese human twins[10] and the dataset of a lignocellulose-degrading community from within a cow rumen[15].

### 4.2.1   Human gut dataset

The contigs from both samples, TS28 and TS29, were pooled. In the same way, scaffolds from TS28 and TS29 were pooled. All scaffolds were longer than 1000 bp (Table 4).

**Table 4: Properties of the real human gut dataset.**

| FASTA file | Sequences | Mb |
|---|---|---|
| Contigs | 153,564 | 255.2 |
| Contigs ≥ 1000 bp | 63,399 | 187.1 |
| Scaffolds | 18,172 | 164.4 |

### *4.2.2   Cow rumen dataset*

The same dataset as in Dröge *et al.*[9] was used. As the scaffolds of the assembled contigs were of lower quality than the contigs (A. Sczyrba, pers. comm.), scaffolds were split into contigs at all gaps consisting of at least 200 "N" characters. We subsequently split the resulting contigs of at least 10 kb into 'chunks' of 2000 bp, resulting in at least five chunks for each contig (Table 5).

**Table 5: Properties of the chunked cow rumen dataset.**

| FASTA file | Sequences | Mb |
|---|---|---|
| Contigs | 159,263 | 318.5 |
| Scaffolds | 12,192 | 369.4 |

## 4.3   Reference data

The NCBI taxonomy,[54] downloaded on 11/22/2012, was used as the reference taxonomy. The following reference databases from the NCBI were pooled to generate our reference sequence (RS) collection: NCBI genomes (downloaded on 11/22/2012), NCBI draft bacterial genomes (downloaded on 11/22/2012), the NCBI human microbiome project (downloaded on 10/16/2012) and NCBI RefSeq[55] microbial version 56. Subsequently, duplicate sequences were removed to make the RS collection non-redundant. This RS collection contained sequences for 841 different genera, 2543 different species and 4516 different strains. The total size of the RS collection was 16 Gb.

In the marker gene (MG) analysis, the following MG sequence collections and HMM profiles were used: For the 16S and 23S MG analysis, bacterial and archaeal reference sequences from the SILVA database[56] were retrieved (version 111, released on 7/27/2012). The corresponding taxonomic identifiers were mapped onto the NCBI taxonomy. The resulting collection contained 126,742 sequences for 39,201 different species (199 Mb in total).

For the 5S MG analysis, MG sequences were retrieved from NCBI on 2/8/2013 via Maglott *et al.*[57]; the collection contained 12,424 sequences for 1278 species (5.8 Mb in total).

In addition, reference sequences for the following 31 bacterial marker gene families were retrieved from NCBI on 2/8/2013 via Maglott *et al.*[57]: *dnaG, infC, pgk, rpoB, tsf, frr, nusA, pyrG, rpmA, smpB, rpsC, rpsI, rpsK, rpsS, rpsB, rpsE, rpsJ, rpsM, rplA, rplB, rplC, rplD, rplE, rplF, rplK, rplL, rplM, rplN, rplP, rplS* and *rplT*. This MG collection contained 63,530 sequences for 1380 different species (52 Mb in total).

HMM profiles for the 16S, 23S, and 5S marker genes were retrieved from Huang *et al.*[40] HMM profiles trained on the protein families for the 31 bacterial MG were retrieved from Wu & Scott.[7]

## 4.4 Test environments

The benchmarks were run on different hardware configurations. When measuring runtime, Hardware Configurations 1 or 2 were used if not stated otherwise.

1. Server: AMD Opteron Processor 6386 SE, 2.8 GHz; 512 GB RAM; local SSD storage; Debian GNU/Linux 7.1.
2. Laptop: Intel i5 M520 2.4 GHz; 4 GB RAM; 7200 rpm laptop storage; Windows 7 64-bit, Ubuntu 12.04 64-bit; Oracle VirtualBox 4.2.12: 2 GB RAM, 8 GB swap, 100 GB HDD, Ubuntu 12.04 64-bit.
3. Server: Intel Xeon CPU X5660, 2.8 GHz; 73 GB RAM; network storage; Debian GNU/Linux 6.0.7.
4. Server: AMD Opteron Processor 6174, 2.2 GHz; 100 GB RAM; local storage; Debian GNU/Linux 6.0.7.
5. Laptop: Intel i5 2557M 1.7 GHz; 4GB RAM, SSD storage, OS X 10.7.

## 4.5 MEGAN4 configuration

*NCBI BLAST* (2.2.27+) was used to generate alignments (Section 4.4, HW Configuration 1), using 15 threads; the tabbed output format (7) was used. *MEGAN4* (4.70.4)[8] was used for taxonomic assignment on a laptop (Section 4.4, HW Configuration 2) using the following settings: *minsupport=5, minscore=2, toppercent=20, mincomplexity=0.44*. The runtime of *MEGAN4* was just a few seconds, as the *LCA* algorithm it uses is simple and fast. Construction of the *BLAST* database from the reference sequence collection required 6 h 55 m, with the size of the database being 4 GB. To simulate the new strain, species and genus scenarios (Table 9: Scenarios 5, 8 and 9), the corresponding alignments of sequences present in both the test and reference data were removed from the *BLAST* output (Table 6).

**Table 6: Runtimes of *BLAST* for the different metagenome datasets.**

| Dataset | Runtime |
|---|---|
| Simulated uniform | 52 m 11 s |
| Simulated log-normal | 19 m 18 s |
| Chunked cow rumen (contigs) | 43 m 29 s |
| Chunked cow rumen (scaffolds) | 42 m 56 s |
| Human gut (contigs) | 44 m 05 s |
| Human gut (scaffolds) | 25 m 37 s |

## 4.6   Taxator-tk configuration

*LAST* (287)[58] was used to produce alignments using one thread, output format 1 (maf). Constructing the *LAST* database for the reference sequence database required 81 h 29 min. The size of the resulting database was 91 Gb (Table 7, Section 4.4, HW Configurations 1, 4).

**Table 7: Runtimes of *LAST* for the different metagenome datasets.**

| Dataset | Runtime (HC 1) | Runtime (HC 4) |
|---|---|---|
| Simulated uniform | 9 h 56 m 27 s | 12 h 10 m 57 s |
| Simulated log-normal | 5 h 02 m 03 s | 6 h 16 m 02 s |
| Chunked cow rumen (contigs) | 12 h 23 m 29 s | 15 h 39 m 24 s |
| Chunked cow rumen (scaffolds) | 15 h 15 m 20 s | 19 h 15 m 12 s |
| Human gut (contigs) | 10 h 29 m 12 s | 13 h 48 m 57 s |
| Human gut (scaffolds) | 7 h 41 m 05 s | 10 h 16 m 20 s |

*Taxator-tk*[9] was then employed to process metagenome sequence fragments using 15 threads and to produce taxonomic assignments using one thread for the input sequences (Table 8, Section 4.4, HW Configuration 4). For the simulated datasets, the corresponding alignments of sequences present in both the test and reference data were removed to simulate the new strain, species and genus scenarios (Table 9: Scenarios 5, 8 and 9).

**Table 8: Runtimes of *taxator-tk* for different metagenome datasets.**

| Dataset | Process fragments | Bin |
|---|---:|---:|
| Simulated uniform | 36 h 54 m 02 s | 17.4 s |
| Simulated uniform (new strain) | 8 h 53 m 20 s | 18.2 s |
| Simulated uniform (new species) | 4 h 44 m 27 s | 18.1 s |
| Simulated uniform (new genus) | 54 m 39 s | 17.5 s |
| Simulated log-normal | 25 h 25 m 49 s | 16.8 s |
| Simulated log-normal (new strain) | 3 h 09 m 16 s | 17.9 s |
| Simulated log-normal (new species) | 2 h 06 m 29 s | 17.4 s |
| Simulated log-normal (new genus) | 36 m 34 s | 16.9 s |
| Chunked cow rumen (contigs) | 3 h 03 m 07 s | 24.9 s |
| Chunked cow rumen (scaffolds) | 46 m 59 s | 19.2 s |
| Human gut (contigs) | 6 h 38 m 56 s | 22.5 s |
| Human gut (scaffolds) | 2 h 47 m 50 s | 18.6 s |

**Commands**

*LAST command:*

lastal -f 1 lastDb query.fna | lastmaf2alignments.py | sort | gzip > alignments.gz

*BLAST command:*

blastn -db blastDb -query query.fna -num_threads 15 -outfmt '6 qseqid qstart qend qlen sseqid sstart send bitscore evalue nident length' -out alignments.blast

*Produce fragments:*

cat alignments.blast | alignments-filter -b 50 | taxator -a rpa -q query.fna -f ref.fna -g ref_all.tax -p 15 | sort > fragments.gff3

*Produce assignments:*

cat fragments.gff3 | binner > assignments.tax

## 4.7 PPS+ configuration

*PPS+* benchmarks were run using one thread (Section 4.4, HW Configuration 3). The *PPS+* configuration file specifies the default values of the parameters used.

## 4.8 PPS generic model configuration

*PPS* was run using one thread (Section 4.4, HW Configuration 3). *PPS* was trained to include the 200 most abundant genera in the reference sequences (Section 4.3). The *PPS* models were built down to the genus rank, as this is the default setting of *PPS*.

## 4.9 Assignment quality measures

### 4.9.1 Micro-averaged precision and recall

To assess the quality of the taxonomic assignments for the simulated datasets, we evaluated the micro-averaged precision (sometimes also known as the micro-averaged specificity) and the micro-averaged recall (sometimes also known as the micro-averaged sensitivity) of taxonomic assignments for the different methods, as detailed below. Both measures were calculated based on the number of assigned bp for each taxonomic rank, instead of per assigned fragment, as the correct assignment of larger sequence fragments is more beneficial for the retrieval of "draft genome" bins than for short fragments.

The micro-averaged precision was defined as:

$$p^l = \frac{\sum_{i=1}^{N_p^l} TP_i^l}{\sum_{i=1}^{N_p^l} TP_i^l + FP_i^l};$$

and micro-averaged recall was defined as:

$$r^l = \frac{\sum_{i=1}^{N_r^l} TP_i^l}{\sum_{i=1}^{N_r^l} TP_i^l + FN_i^l};$$

where $l$ denotes the taxonomic rank evaluated, such as species, genus, family, order, class, phylum or superkingdom; $(TP_i^l + FN_i^l)$ is the number of bp from taxon $i$; $(TP_i^l + FP_i^l)$ is the number of bp assigned to taxon $i$ and $TP_i^l$ is the number of bp correctly assigned to taxon $i$. The precision is micro-averaged over all bins $N_p^l$ to which a sequence fragment was assigned and the recall is micro-averaged over all $N_r^l$ taxa present in the simulated dataset at rank $l$.

The micro-averaged precision is the fraction of correctly assigned bp from all predictions for a particular taxonomic rank and represents a measure of confidence for the predictions of a method. The micro-averaged recall is the fraction of correct assignments of the test sample for a particular taxonomic rank. To avoid an uninformative increase of the micro-averaged recall by having unassigned sequences which belong to no taxon at a given rank, our test datasets were generated from sequenced isolates with taxa defined at all major taxonomic ranks. Note that for simplification, we denoted the micro-averaged precision as 'precision' and the micro-averaged recall as 'recall' in this document.

### *4.9.2 Taxonomic assignment correction for assessment of bin quality*

Often, a species within a metagenome sample is not directly represented among the reference sequences; however, this respective species is closely related to a species for which there is enough data in the RS or MG collections. In this case, the species from the sample may be consistently assigned to the closely related species. This error does not impact draft genome reconstruction in terms of reconstructing a bin as a set of sequences originating from the same sample population, but the assigned identifier itself is incorrect. To quantify the binning performance independently from taxonomic label assignment, we applied a correction procedure and re-computed the corrected micro-averaged precision and recall values: If most of the sequences (i.e. at least *(correctLabelThreshold \* 100)% bp*) from one taxon were consistently assigned to a false identifier, their identifiers were changed to the correct one, and micro-averaged precision and recall were re-computed. The default setting for the configuration parameter *correctLabelThreshold* was 0.9. The micro-averaged precision and recall were always calculated with and without this correction.

## 4.10 Scaffold-contig consistency definitions

### *4.10.1 Comparison of scaffold and contig assignments*

To assess the consistency of scaffold and contig assignments for a metagenome sample, we define the following measures at all major taxonomic ranks (i.e. superkingdom, phylum, class, order, family, genus and species).

Let us assume that a metagenome sample consists of $m$ scaffolds $s_0, \dots, s_{m-1}$ and $n$ contigs $c_0, \dots, c_{n-1}$, where scaffold $s_k$ consists of $n_k$ contigs $c_{k(0)}, \dots, c_{k(n_k-1)}$. Let function $l$ denotes the taxonomic identifier of a contig or a scaffold at the taxonomic rank being considered, i.e. $l(c_i)$ is a label of the $i^{\text{th}}$ contig and $l(s_k)$ is the label of the $k^{\text{th}}$ scaffold. The lengths of contig $c_i$ and scaffold $s_k$ are denoted by $len(c_i)$ and $len(s_k)$, respectively. Now, we can define the consistency measures 'kb agreement' (Def. 0a) and '% agreement' (Def. 0b) as:

0a) 'kb agreement':
$$a_{kb} = \sum_{k=0}^{m-1} \sum_{j \in \{k(0),\dots,k(n_k-1)\},\ l(s_k)\ and\ l(c_j)\ defined,\ l(s_k)=l(c_j)} len(c_j);$$

0b) '% agreement':

$$a_\% = \frac{a_{kb}}{\sum_{j=0}^{n-1} len(cj)}.$$

In other words, in 'kb agreement' (Def. 0a), the index $k$ goes over all scaffolds, the index $j$ goes over all contigs within a corresponding scaffold. If both labels of scaffold $k$ and contig $j$ are defined and assigned to the same taxa, then the length of contig $j$ is added to the overall sum of lengths of consistently assigned contigs.

### 4.10.2 Taxonomic scaffold-contig assignment consistency

To provide more detailed insights into the evaluation of the binning results of real metagenome datasets, we introduced new detailed measures of the scaffold-contig consistency (described below).

We assume that all contigs $c_0,...,c_{n-1}$ of a particular scaffold originated from the same organism and thus should be assigned the same taxonomic identifier. Let us denote an identifier of contig $c_i$ as $l_i$. Each path $p_i$ from the root of the taxonomy to identifier $l_i$ represents a hypothesis about the identifier of the whole scaffold. We base our definition on the assumption that the most representative identifier of a scaffold corresponds to the path to which the identifiers of all taxonomically assigned contigs that do not lie on the path have the shortest collective weighted distance. Note that we do not have to consider the path $p_i$ from the root to $l_i$ as a potential taxonomic identifier if there is a path $p_j$ from the root to the taxonomic identifier $l_j$ of another contig $c_j$ for which $l_i$ lies on $p_j$ and $i \neq j$, as the shortest collective weighted distance of all contigs of a scaffold to path $p_j$ is always lower than the collective weighted distance to path $p_i$. Let us denote the length of contig $c_i$ as $|c_i|$ (counted in bp). Let us define the weight of contig $c_i$ as $w_i = \frac{|c_i|}{\sum_{j=0}^{n-1} |c_j|}$. Let $tax\_dist(l_i, p_j)$ be the taxonomic distance (i.e. the number of edges in the reference taxonomy) between identifier $l_i$ and the closest identifier $l_k$ that lies on path $p_j$ (This is simply the distance between identifier $l_k$ and path $p_j$). The weighted distance from path $p_j$ to all other identifiers $l_i$ is defined as: $dist(p_j) = \sum_{i=0}^{n-1} w_i * tax\_dist(l_i, p_j)$. Let $p_k$ be the path with the minimum weighted distance ($dist$) from all other identifiers. All contigs $c_i$ that lie on path $p_k$ are considered to be consistently assigned within the scaffold; all contigs $c_j$ that do not lie on the path are considered to be inconsistent. The consistency of the scaffold is then defined as:

1) Proportion of consistently assigned contigs:

$$\frac{|\{c_i \mid l_i \text{ on } p_k\}|}{|\{c_i \mid i=0\ldots n-1\}|},$$

2) Proportion of consistent contigs in bp:

$$\frac{\sum_{\{i \mid l_i \text{ on } p_k\}} |c_i|}{\sum_{i=0}^{n-1} |c_i|},$$

3) Average distance to the path:

$$\frac{\sum_{i=0}^{n-1} tax\_dist(l_i, \ p_k)}{n},$$

4) Average weighted distance to the path:

$$dist(p_k);$$

5) Average distance to the scaffold identifier:

$$\frac{\sum_{i=0}^{n-1} tax\_dist(l_i, \ l_k)}{n},$$

6) Average weighted distance to the scaffold identifier:

$$\sum_{i=0}^{n-1} w_i * tax\_dist(l_i, l_k).$$

The first definition is the coarsest measure and the last is the finest for taxonomic assignment consistency.

We can also group the scaffolds using $l_k$ and compute the measures for individual taxa. However, these groups do not correspond to the assigned bins, as a scaffold's taxonomic identifier does not always correspond to the taxonomic identifier of the lowest assigned contig of that scaffold.

The consistency of the entire sample can also be defined as the (weighted) average of these measures. Let $s_0, \ldots, s_{m-1}$ be all scaffolds in the sample, where if a contig is not assigned to a scaffold, an artificial scaffold that contains this one contig is created. We can also consider only scaffolds that contain only a certain number of contigs or those that are at least *x* bp long, for example.

Thus if we compute these measures for two different binning methods, we can assess the consistency of the respective taxonomic assignments at six different levels. However, be aware that it is recommended to also look at the number of bp assigned at different taxonomic ranks by each method, since the consistency of a method that assigns everything to the root of the taxonomy seems to be perfect according to these scaffold-contig consistency definitions.

# 5 Results

We evaluated *PPS+* by comparing it to homology-based methods (*MEGAN4*, *taxator-tk*), the composition-based method *PhyloPythia* trained under expert guidance (a recommended but time-consuming procedure) and to a generic *PPS* model using default settings for all methods (Sections 4.5–4.8). For *PPS*, we generated a generic model for the 200 most abundant genera represented in the reference sequence collection, which currently includes 2543 species (Section 4.3). As benchmark datasets, we created two simulated datasets – one with a uniform (137 Mb) and one with a log-normal (66 Mb) distribution of 47 community members (Section 4.1) – and used two real datasets, a pooled metagenome sample from the guts of two obese human twins[10] and a cow rumen metagenome sample from Hess *et al.*[15] As in Dröge *et al.*[9], every scaffold of the cow rumen dataset was split into contigs at all stretches of at least 200 "N" characters and the resulting contigs of at least 10 kb were split into 2-kb fragments ('chunks'). In the evaluation of this dataset, we considered the contigs of at least 10 kb as scaffolds and chunks of 2 kb as contigs (Section 4.2).

## 5.1 Benchmarks with simulated datasets

We constructed the simulated datasets by assembling simulated paired-end 90-bp reads with an empirical error profile, to make them as realistic as possible (Section 4.1). For the evaluation of the taxonomic classifiers on the simulated datasets, micro-averaged precision and recall were calculated (Section 4.9). To assess the performance of the different methods in assigning sequence fragments from microbial community members without related sequence reference genomes being available, 'new strain', 'new species' and 'new genus' scenarios were simulated by removing all sequence data from the taxa of the simulated test dataset at each rank from the reference sequence collections and, optionally, from the marker gene sequence collection. For instance, in the new species scenario, all reference sequences of the same species as the simulated community members were excluded as reference data for *PPS* model creation. Furthermore, for *PPS+*, we distinguished whether the reference data were excluded (masked) from the reference sequence (RS) collection or also from the marker gene (MG) collection, since the MG collection included sequences for 15 times as many distinct species than the RS collection and these were therefore two different situations to consider. Micro-averaged precision and recall were computed for all nine scenarios (Table 9). Furthermore, these measures were also calculated with a 'correction', to account for the case where the sequences of one taxon were consistently assigned to a different taxon, because for draft genome reconstruction, it is more important that the sequences are assigned consistently than that the taxonomic identifier is correct.

**Table 9: Test scenarios.**

Test scenarios where data was removed (masked) up to the specified rank for the corresponding taxa represented in the simulated metagenome datasets from the reference collections. RS denotes the reference collection of complete or draft genomes; MG indicates the reference collection of marker genes (Section 4.3).

| Test scenario | Rank masked from RS | Rank masked from MG |
|---|---|---|
| 1. | None | None |
| 2. | Strain | None |
| 3. | Species | None |
| 4. | Genus | None |
| 5. | Strain | Strain |
| 6. | Species | Strain |
| 7. | Genus | Strain |
| 8. | Species | Species |
| 9. | Genus | Genus |

*PPS+* showed a substantially improved micro-averaged precision and recall over the *PPS* generic model, which demonstrated the impact of the improved selection of training data and modeled taxa (Figs 1, 5a–5d and 7a–7d). *PPS+* almost always had higher micro-averaged precision and recall than *MEGAN4,* except when almost all test data were included in the reference sequences (Figs 1A, 1C, 5a–5c, 5e, 7a–7c and 7e). This was even more pronounced when comparing bin quality using the corrected measures (Figs 1B, 1D, 6a–6c, 6e, 8a–8c and 8e). When comparing *PPS+* to *taxator-tk*, *PPS+* had substantially improved micro-averaged recall, particularly for lower ranks (Figs 1A, 5a–5c, 5f, 7a–7c and 7f); while *taxator-tk* outperformed all other methods in terms of micro-averaged precision (Figs 1A, 5a–5f and 7a–7f). Both methods were similarly precise when analyzing bin recovery, independent of assigning the taxonomic identifiers to the corrected measures (Figs 1B, 1D, 6a–6c, 6f, 8a–8c and 8f). A strong point of *PPS+* is that it more rarely predicted wrong taxa that were not a part of the metagenome sample in comparison to the other methods (Figure 9). For example, for the genus rank in Scenarios 3 and 8, *PPS+* assigned sequences to only 2–5 false positive taxa, while *taxator-tk* identified 20, *MEGAN4* 37 and the *PPS* default model 59. If *PPS+* identified wrong taxa, these were usually very closely related to the true taxa.

**Figure 5: Benchmark results for the simulated dataset with uniform distribution.**

Micro-averaged precision (P) and recall (R) (Section 4.9.1) at different taxonomic ranks were calculated for (panels *a–c*) *PPS+*, (panel *d*) the generic *PPS* model, (panel *e*) *MEGAN4* and (panel *f*) *taxator-tk* in all test scenarios (Table 9: Test Scenarios 1–9). In parentheses, (mg) and (rs) denote whether the sequences at a given taxonomic rank were masked from the marker gene or from the reference sequence collections, respectively (Sections 4.1 and 4.3). If not stated, sequences were masked from both reference collections.

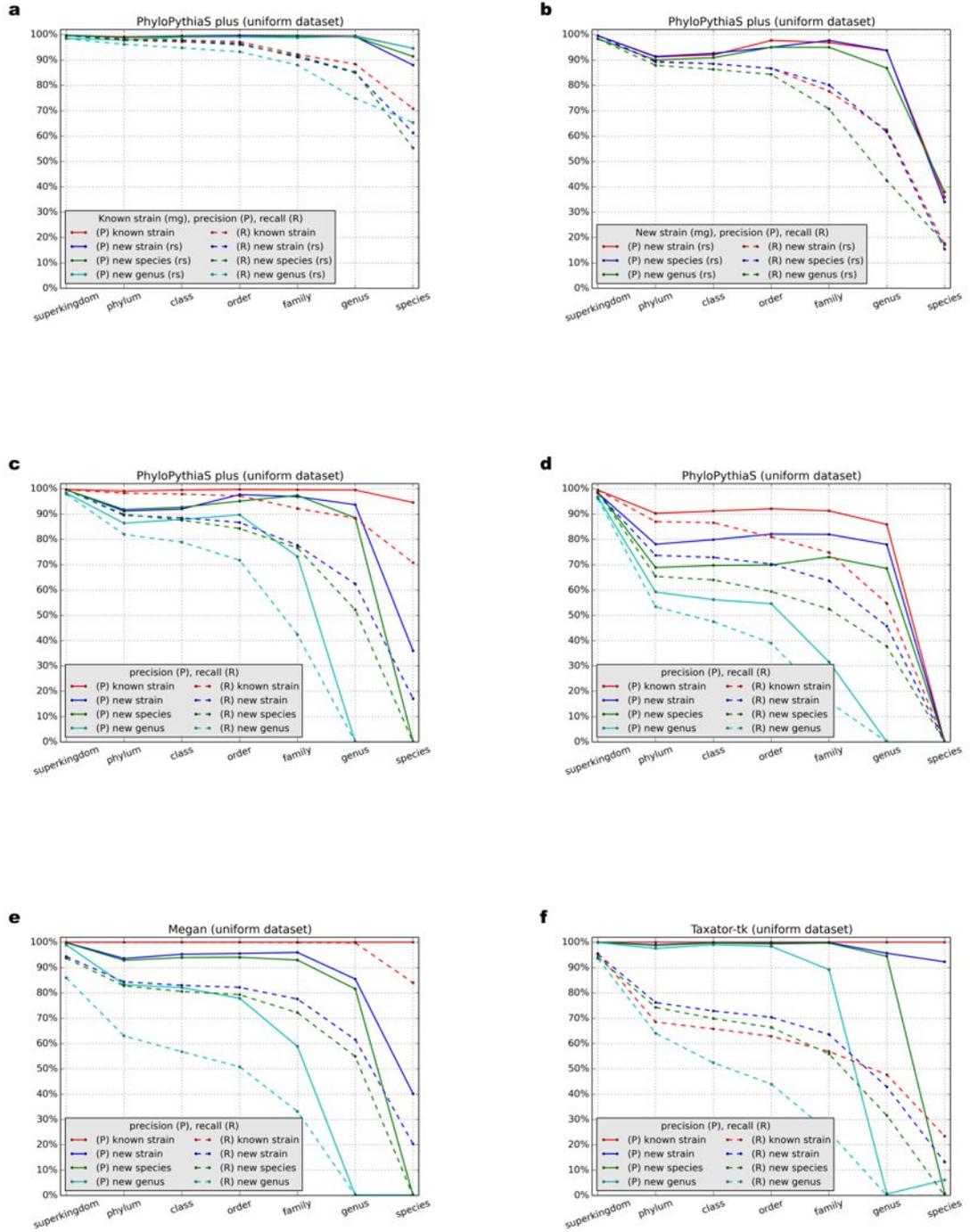

**Figure 6: Benchmark results for the simulated dataset with uniform distribution using 'correction'.**

Micro-averaged precision (P) and recall (R) were calculated with a 'correction' (Section 4.9) at different taxonomic ranks for (panels *a–c*) *PPS+*, (panel *d*) the generic *PPS* model, (panel *e*) *MEGAN4* and (panel *f*) *taxator-tk* in all test scenarios (Table 9: Test Scenarios 1–9, Section 4.1).

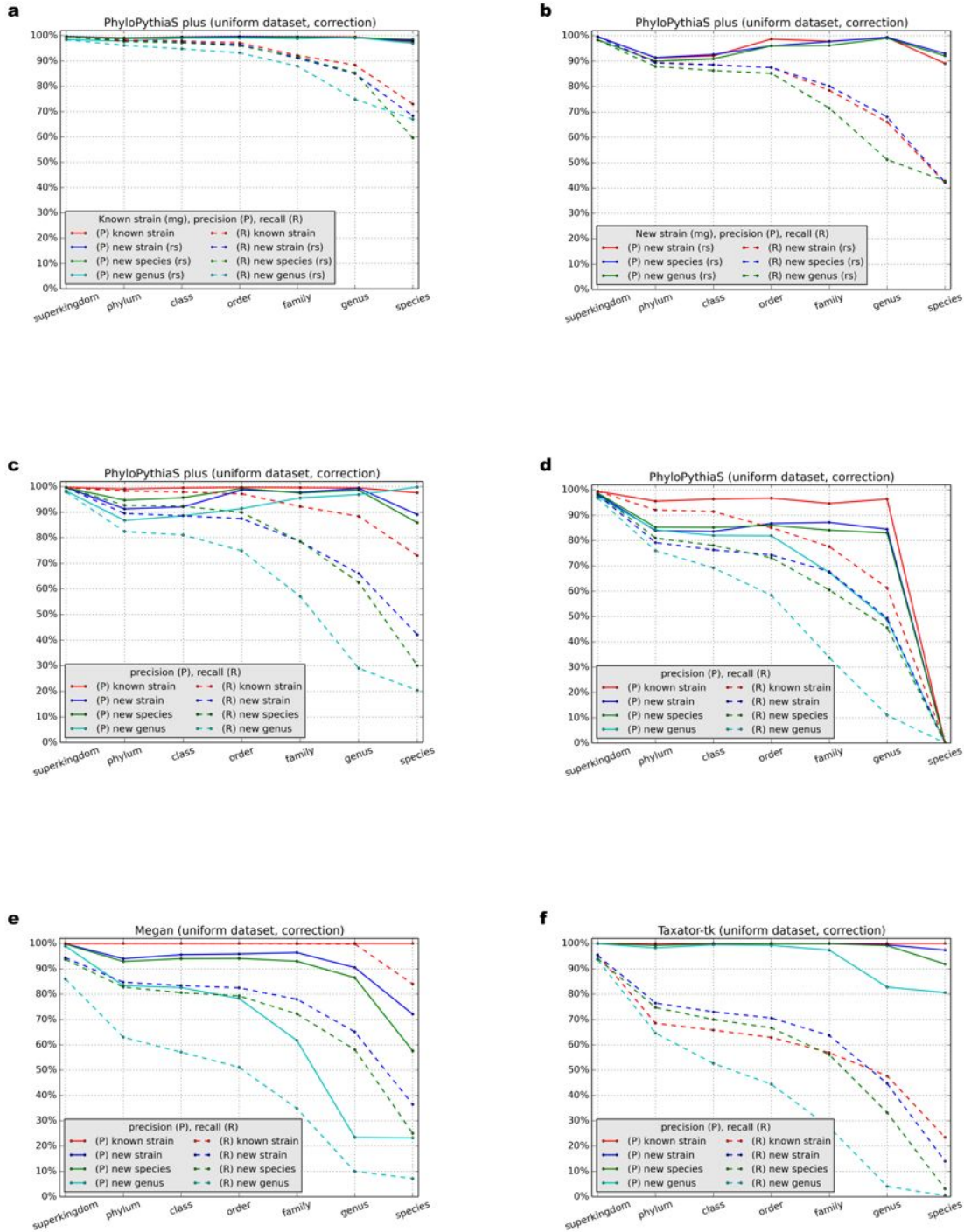

**Figure 7: Benchmark results for the simulated dataset with the log-normal distribution.**

Micro-averaged precision (P) and recall (R) (Section 4.9.1) at different taxonomic ranks were calculated for (panels *a–c*) *PPS+*, (panel *d*) the generic *PPS* model, (panel *e*) *MEGAN4* and (panel *f*) *taxator-tk* in all test scenarios (Table 9: Test Scenarios 1–9, Section 4.1).

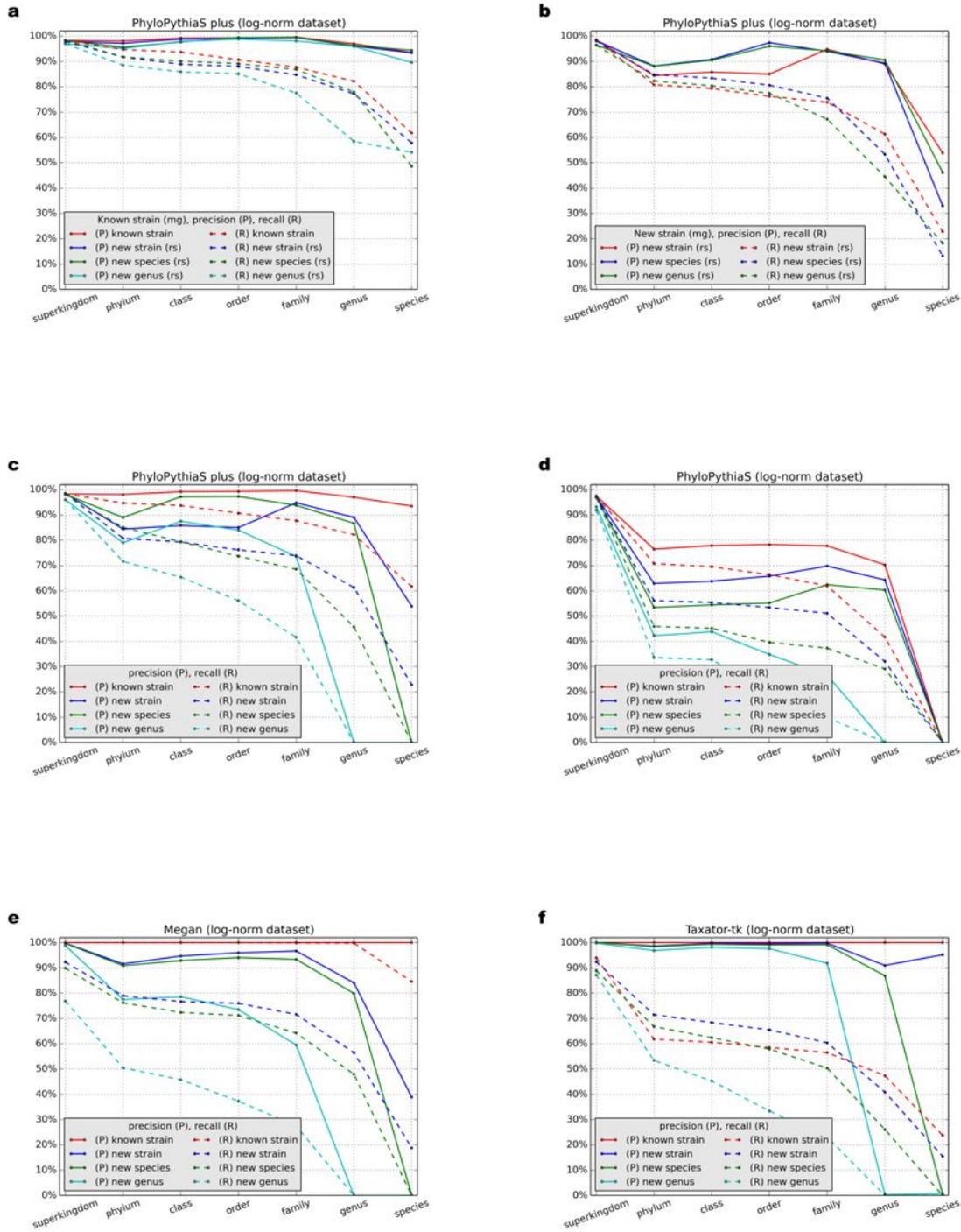

**Figure 8: Benchmark results for the simulated dataset with the log-normal distribution using 'correction'.**

Micro-averaged precision (P) and recall (R) were calculated with a 'correction' (Section 4.9) at different taxonomic ranks for (panels *a–c*) *PPS+*, (panel *d*) the generic *PPS* model, (panel *e*) *MEGAN4* and (panel *f*) *taxator-tk* in all test scenarios (Table 9: Test Scenarios 1–9, Section 4.1).

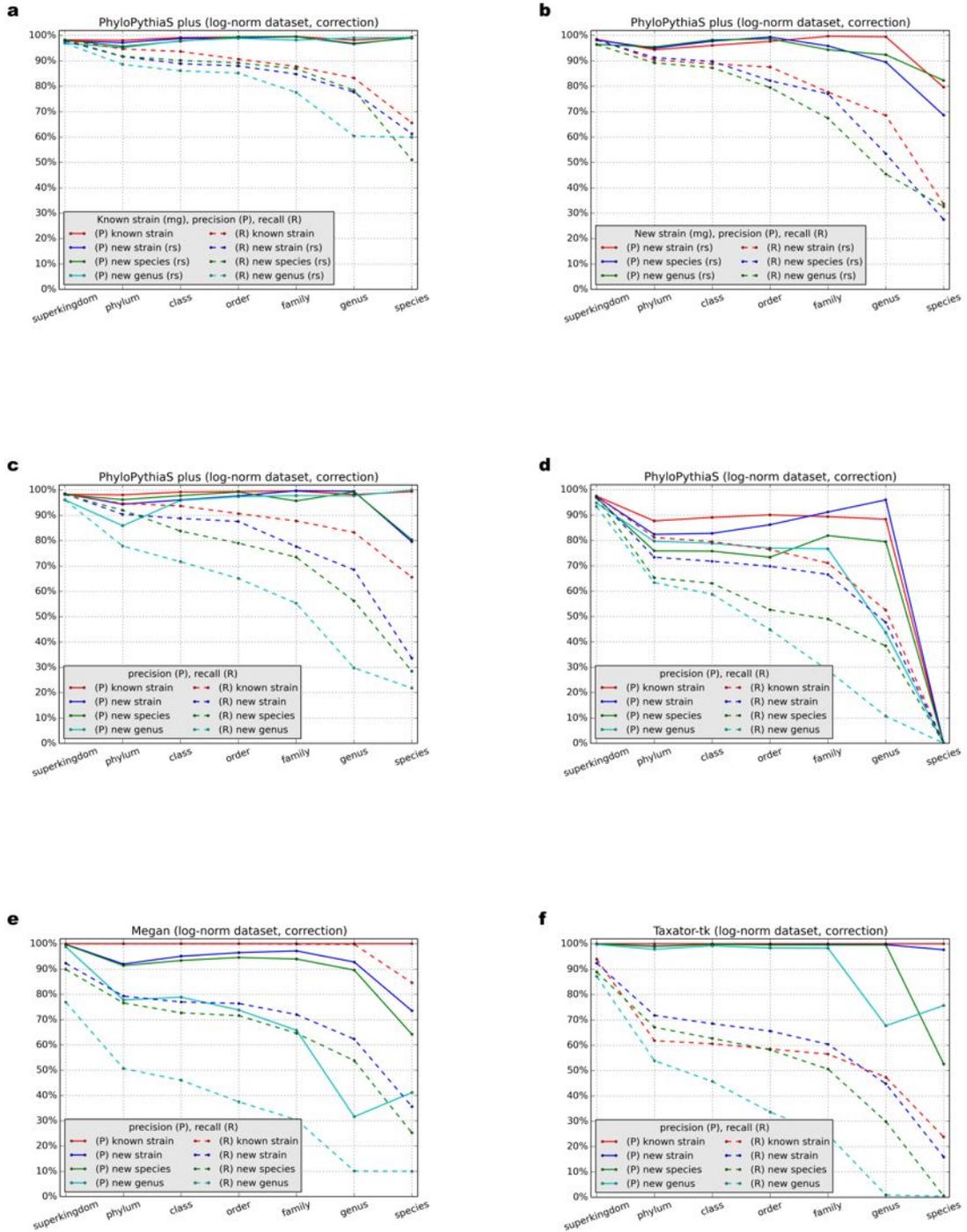

**Figure 9: Base pairs assigned to individual taxa for a simulated metagenome of a microbial community with log-normally distributed species abundance.**

The number of taxonomic assignments to each taxon in bp is indicated on a log-scale by the pie chart sizes for *PPS+* (red), the generic *PPS* model (purple), *taxator-tk* (blue), *MEGAN4* (yellow) and the underlying standard of truth (black). There were 47 strains present in the simulated metagenome sample. Assignments to taxa not shown in black in the chart are to false taxa that are not present in the simulated metagenome. Panel *a* shows the scenario where sequences from the same species as those of the simulated dataset were excluded from the reference sequences but not the marker gene databases (Table 9: Test Scenario 3). Panel *b* shows the scenario where sequences from the same species as those of the simulated dataset were excluded from the reference sequence and marker gene databases (Table 9: Test Scenario 8).

**a**

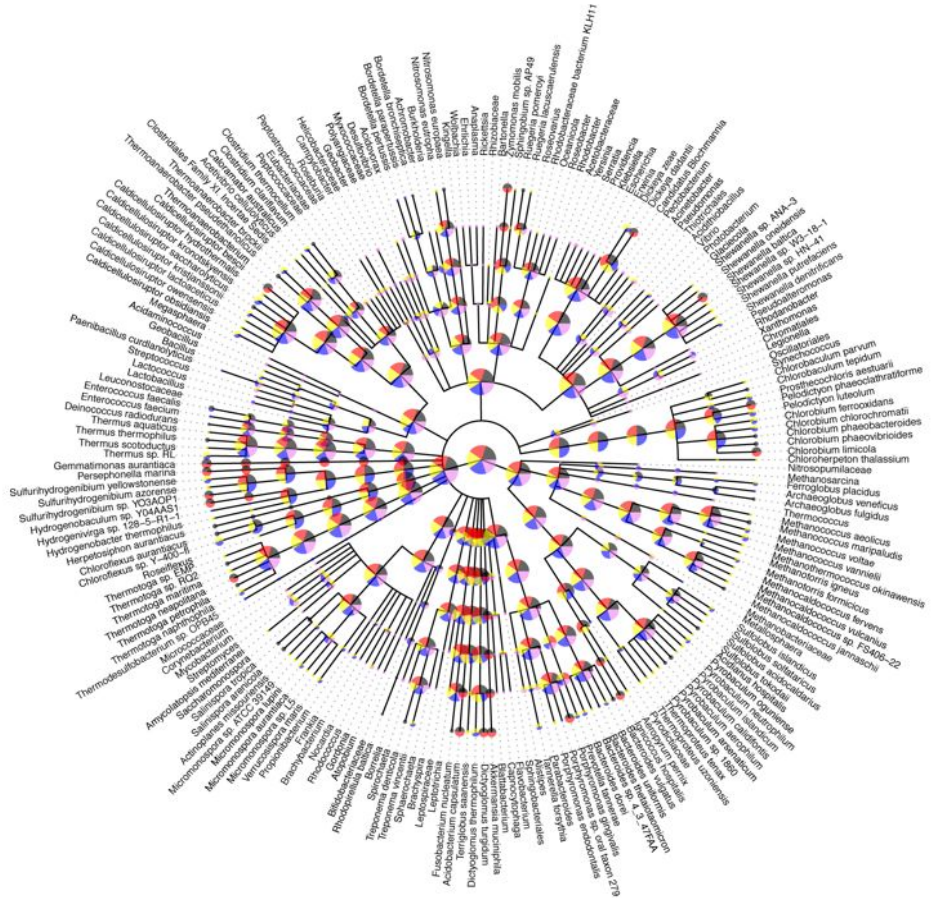

**b**

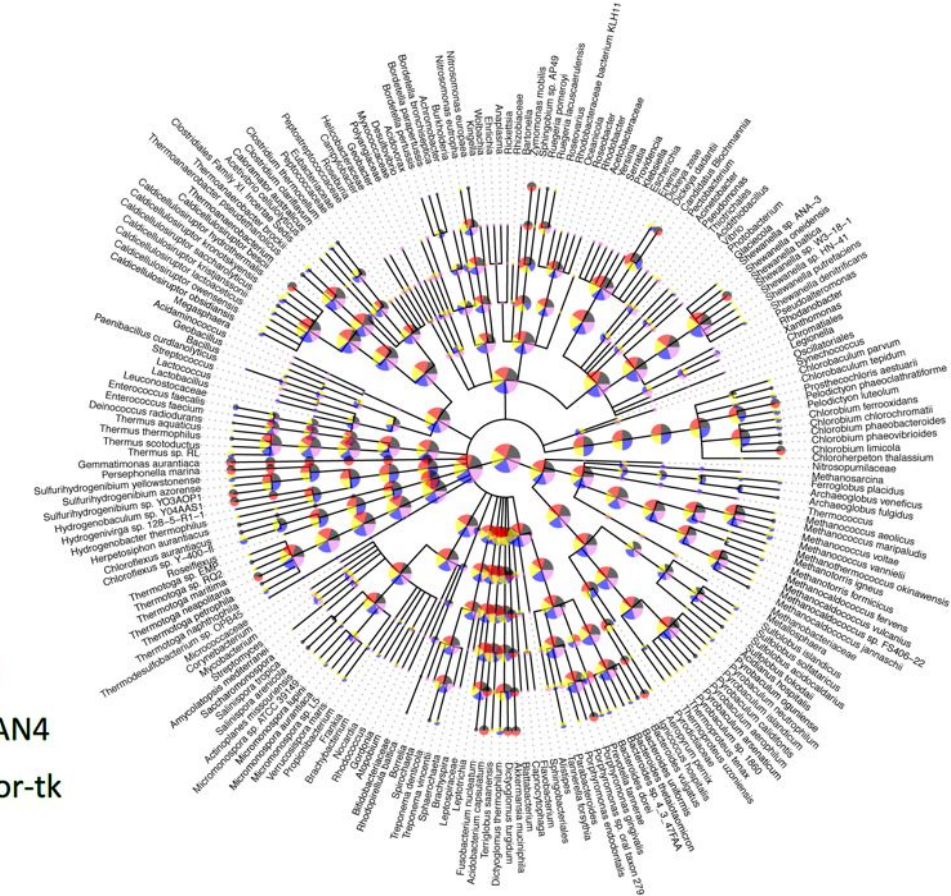

- ■ TRUE
- ■ MEGAN4
- ■ taxator-tk
- ■ PPS
- ■ PPS+

## 5.2 Detailed results for the simulated data sets

This section provides a detailed description of the results of the benchmarks with simulated datasets in nine different test scenarios (Table 9). *PPS+*, *PPS* generic model, *MEGAN4* and *taxator-tk* were compared to each other in terms of micro-averaged precision and recall (Section 4.9). The nine different scenarios evaluate assignment performances for different evolutionary distances between the sample sequences and the available reference sequences. For instance, in (Table 9: Test Scenarios 6), all sequences from the species included in the simulated communities were excluded from the reference sequence collection and all sequences of the same strains were excluded from the marker gene sequence collection.

### 5.2.1 Uniform dataset

For *PPS+*, a drop in both micro-averaged precision and recall was only observed for low-level taxonomic assignments when removing reference data from the same strain, species or genera from the shotgun (RS) collection and also from the MG collection (Table 9: Test Scenarios 2, 3 and 4 versus Test Scenarios 5, 8 and 9), which demonstrated that for microbial community members that have been profiled by 16S sequencing but which have no sequenced genomes available, *PPS+* can perform highly accurate low-level taxonomic assignments, unlike from all other tested methods (Figs 5a and 5c–5f).

In more detail, *PPS+* showed substantially higher micro-averaged precision and recall than the *PPS* generic model for all test scenarios (Fig. 5a–5d, Table 9: Test Scenarios 1–9). *PPS+* also showed substantially higher micro-averaged precision and recall than *MEGAN4* for the assignment of sequences from new strains, species and genera (Figs 5a and 5e, Table 9: Test Scenarios 2–4), when these were represented in the reference collection as marker genes. An exception was the unrealistic case, when all of the simulated metagenome data were available in the reference sequence collection (Table 9: Test Scenario 1).

Simulating the situation where the microbial community members have not been observed in profiling before, we removed these strains from the MG collection and the shotgun data (RS) for the strains, species or genera of the simulated metagenome datasets (Table 9: Test Scenarios 5, 6 and 7). We removed more data from the shotgun sequence (RS) collection than from the MG collection to simulate the situation where a closer relative can be found among the marker genes and a more distant one among the sequenced genomes, as many taxa have been profiled but have not had their genomes sequenced. *PPS+* assignment quality (both micro-averaged precision and recall) dropped in comparison to the situation where strains have been profiled (Fig. 5a,b). However, it was still better than *MEGAN4* (Figure 5e) for all ranks, except for the lowest-level assignment (species), when the strains were removed from the RS collection only (Table 9: Test Scenario 5). As the removal of strain-level data in many cases also removed all data for the respective species from the RS collection, both methods made false assignments to related species in these scenarios.

When we removed even more reference data from the MG collection to simulate the binning of microbial community members for which no members of the same species or genera have been profiled or sequenced before (Figure 5c, Table 9: Test Scenarios 8 and 9), the micro-averaged precision for ranks above remained high (Table 9: Test Scenario 8, genus rank: 88.5%; Test Scenario 9, family rank: 73.2%), while the micro-averaged recall dropped moderately. However, *PPS+*'s assignments were still substantially better than those of *MEGAN4* for these ranks (Figure 5e, Test Scenario 8, genus rank: 81.6%; Test Scenario 9, family rank: 58.9%). For lower ranks for which all reference data were removed, both methods had low micro-averaged precision and recall due to false positive assignments.

*Taxator-tk* showed a lower micro-averaged recall than *PPS+* across all tested scenarios (Figs 5a–5c and 5f), but showed outstanding micro-averaged precision for the order rank and above (close to 100%), and never dropped below 89% at lower ranks. Thus this method could also be used for taxonomic profiling to determine the presence of particular taxa reliably in a given sample.

### 5.2.2   Log-normal dataset

Even though the log-normal dataset was more challenging for all the tools, this benchmark yielded similar conclusions as the benchmark with the uniform dataset.

*PPS+* performed substantially better than the generic *PPS* model in terms of the micro-averaged precision and recall in all test scenarios (Fig. 7a–7d, Table 9: Test Scenarios 1–9).

At low taxonomic ranks (i.e. family, genus and species), *PPS+* outperformed *MEGAN4* in terms of micro-averaged precision and recall in almost all test scenarios (Figs 7a–7c and 7e, Table 9: Test Scenarios 2–9), except at the family rank in the 'new strain' scenario, where *MEGAN4* had slightly better micro-averaged precision (96.7%) than *PPS+* (94.8%) (Figs 7b, 7c and 7e, Table 9: Test Scenario 5). In the unrealistic case, where all reference data remained in the reference (RS and MG) collections, *MEGAN4* had better micro-averaged precision and recall (Figs 7a–7c and 7e, Table 9: Test Scenario 1).

Overall, *PPS+* showed substantially better micro-averaged recall than *taxator-tk*, whereas *taxator-tk* showed mostly better micro-averaged precision (Figs 7a–7c and 7f, Table 9: Test Scenarios 1–9). Moreover, in the case where microbial community members have been profiled by 16S but have no sequenced genomes, *PPS+* showed a very high micro-averaged precision at low taxonomic ranks (i.e. family, genus and species) 99.5–89.6% (Figs 7a and 7f, Table 9: Test Scenarios 2–4). In several cases, *PPS+* showed better micro-averaged precision than *taxator-tk*; for example, at the family rank, the precision was 98.1% for *PPS+* vs 91.9% for *taxator-tk* (Figs 7a and 7f, Table 9: Test Scenarios 4) and at the genus rank, it was (scenario 2) 96.1%, (scenario 3) 96.3% for *PPS+* vs (scenario 2) 91%, (scenario 3) 86.9% for *taxator-tk* (Figs 7a and 7f, Table 9: Test Scenarios 2, 3).

### 5.2.3 Benchmarks with corrections

In the test scenarios when the reference data were excluded from the MG databases (Table 9: Test Scenarios 5–9), the micro-averaged precision of *PPS+* for low taxonomic ranks (i.e. genus and species) was lower than the micro-averaged precision of *taxator-tk* because of the way *PPS+* chooses the taxa that are modeled. If the sequences from the same strains as those of the simulated metagenome samples were removed from the MG reference database at the strain, species or genus ranks, the marker gene analysis assigned sequences of the metagenome sample that would otherwise have a very good match to the respective MG database sequences to corresponding closely related taxa.

In the subsequent *PPS* training phase, the sample-specific data were used to train closely related clades; moreover, reference sequences from closely related clades were used as training data as well. However, for the draft genome reconstruction, it is necessary to infer consistent bins from a metagenome sample. The actual identifiers of the bins are of lower importance. Therefore, we recomputed the micro-averaged precision and recall measures with a correction to account for the phenomenon described above (Section 4.9, Figs 6a–6f and 8a–8f, Table 9: Test Scenarios 1–9).

The corrected micro-averaged precision of *PPS+* was substantially better than it was without the correction for all scenarios. The difference for the other methods is not that pronounced, since they choose clades to which metagenome sequences are assigned in a different way. When comparing *PPS+* to *MEGAN4* using these corrections, *PPS+* showed higher micro-averaged precision and recall. When comparing *PPS+* to *taxator-tk*, *PPS+* had higher micro-averaged recall; however, neither method was consistently more precise.

## 5.3   Benchmarks with real datasets

As simulated datasets were not likely to be representative for all properties of real datasets, we additionally evaluated the performances of all methods on two real metagenome datasets: an assembled human gut sample generated with a 454 GS FLX Titanium sequencer and an assembled cow rumen sample generated with Illumina GAIIx and Illumina HiSeq 2000 sequencers.

### 5.3.1   *Comparison of scaffold and contig assignments*

For each taxonomic rank, the percentage and the total number of kb (% agreement and kb agreement) that were assigned the same taxonomic identifier were calculated, considering the assignments of the scaffold and contig sequences (Section 4.10.1). Each contig was assigned up to two taxonomic identifiers: one from the contig assignment and the other from the scaffold assignment. These two taxonomic labels were then compared. If we considered contigs with two identical taxonomic labels to be correctly assigned and contigs with two distinct taxonomic labels to be as incorrectly assigned, then "% agreement" resembles a measure of precision (i.e. correctly assigned bp ÷ correctly and incorrectly assigned bp), while "kb agreement" indicates recall (i.e. the total number of correctly assigned bp). For the chunked cow rumen dataset, *taxator-tk* showed the highest assignment consistency (Table 10); however, it assigned much fewer data than the other methods at lower taxonomic ranks. A detailed comparison of the scaffold and contig assignments was performed using heat maps, where the rows correspond to scaffolds and the columns correspond to contig assignments (Fig. 10–17). *PPS+* performed substantially better in terms of both measures (% agreement and kb agreement) than the generic *PPS* model in almost all cases. *PPS+* was also more consistent than *MEGAN4* for all lower ranks and assigned many more sequences than *MEGAN4* overall; for instance, at the genus rank, the scores were *PPS+/MEGAN4:* 84.3/56 (% agreement), 33,724/13,726 (kb agreement). The low number of bp assigned by *MEGAN4* and *taxator-tk* to lower taxonomic ranks reflects the availability of few related reference genome sequences for the cow rumen metagenome sample, which was not an issue for composition-based methods.

For the human gut microbiome, extensive sequencing of isolate cultures has resulted in a large collection of several hundred reference genome sequences. Accordingly, for the human gut dataset, *taxator-tk* and *MEGAN4* assigned many more sequences than they did for the cow rumen dataset (Tables 10 and 11). For *MEGAN4*, this was most pronounced for the genus and species ranks. The most consistent method was again *taxator-tk*, but it also assigned fewer sequences than the other methods. *PPS+* performed better than the generic *PPS* model in all cases in terms of both measures (i.e. % agreement and kb agreement) (Table 11). *PPS+* and *MEGAN4* showed comparable consistency, with *PPS+* being more consistent for the class, order and species ranks, and *MEGAN4* being more consistent for the superkingdom, family and genus ranks. However, *PPS+* assigned (kb agree) more sequences than *MEGAN4*, except for the genus and species ranks. Thus in the case of larger collections of related isolate genome sequences being available, composition- and homology-based methods perform similarly well.

## Table 10: Comparison of contig and scaffold assignments of the chunked cow rumen dataset.

The contigs and scaffolds of the chunked cow rumen dataset were assigned using *PPS+*, the generic *PPS* model, *MEGAN4* and *taxator-tk*. For each method, up to two taxonomic identifiers were assigned to each contig at each rank, i.e. one identifier came from the contig assignment and the second identifier came from the corresponding scaffold assignment. Contigs with less than two taxonomic assignments at each rank were not considered in this comparison. The measure "% agreement" was the percentage of contigs with the same two taxonomic identifiers at a particular rank, whereas "kb agreement" was the total number of kb of contigs with the same taxonomic identifiers (Section 4.10.1). Green numbers correspond to the best values, whereas red numbers indicate the worst values.

| Method | Rank | % agreement | kb agreement |
|---|---|---|---|
| *PPS+* | Superkingdom | 92.3 | 283,950 |
| *PPS* | Superkingdom | 94.6 | 291,643 |
| *MEGAN4* | Superkingdom | 99.6 | 86,402 |
| *taxator-tk* | Superkingdom | 100.0 | 187,292 |
| *PPS+* | Phylum | 73.9 | 153,774 |
| *PPS* | Phylum | 67.8 | 75,538 |
| *MEGAN4* | Phylum | 74.2 | 43,380 |
| *taxator-tk* | Phylum | 98.2 | 59,702 |
| *PPS+* | Class | 86.0 | 99,596 |
| *PPS* | Class | 58.5 | 43,931 |
| *MEGAN4* | Class | 68.5 | 33,780 |
| *taxator-tk* | Class | 97.7 | 23,190 |
| *PPS+* | Order | 88.4 | 98,616 |
| *PPS* | Order | 63.8 | 41,349 |
| *MEGAN4* | Order | 68.9 | 32,650 |
| *taxator-tk* | Order | 98.0 | 22,368 |
| *PPS+* | Family | 80.0 | 46,343 |
| *PPS* | Family | 55.8 | 19,158 |
| *MEGAN4* | Family | 55.0 | 15,790 |
| *taxator-tk* | Family | 98.9 | 7276 |
| *PPS+* | Genus | 84.3 | 33,724 |
| *PPS* | Genus | 63.2 | 12,938 |
| *MEGAN4* | Genus | 56.0 | 13,726 |
| *taxator-tk* | Genus | 99.1 | 6042 |
| *PPS+* | Species | 91.6 | 9821 |
| *PPS* | Species | N/A | N/A |
| *MEGAN4* | Species | 54.6 | 8502 |
| *taxator-tk* | Species | 100.0 | 292 |

**Table 11: Comparison of contig and scaffold assignments of the human gut metagenome dataset.**

Contig and scaffold sequences of the human gut metagenome dataset were assigned using *PPS+*, the generic *PPS* model, *MEGAN4* and *taxator-tk*. The measures "% agreement" and "kb agreement" were used to compare individual methods (Section 4.10.1). Green numbers correspond to the best values, whereas red numbers indicate the worst values.

| Method | Rank | % agreement | kb agreement |
|--------|------|-------------|--------------|
| *PPS+* | Superkingdom | 99.9 | 146,639 |
| *PPS* | Superkingdom | 99.8 | 146,392 |
| *MEGAN4* | Superkingdom | 100.0 | 133,687 |
| *taxator-tk* | Superkingdom | 100.0 | 131,699 |
| *PPS+* | Phylum | 99.0 | 140,283 |
| *PPS* | Phylum | 97.0 | 124,884 |
| *MEGAN4* | Phylum | 99.0 | 127,658 |
| *taxator-tk* | Phylum | 100.0 | 104,475 |
| *PPS+* | Class | 99.5 | 134,707 |
| *PPS* | Class | 96.9 | 118,068 |
| *MEGAN4* | Class | 98.5 | 122,131 |
| *taxator-tk* | Class | 100.0 | 84,228 |
| *PPS+* | Order | 99.5 | 134,127 |
| *PPS* | Order | 97.3 | 117,185 |
| *MEGAN4* | Order | 98.6 | 121,811 |
| *taxator-tk* | Order | 100.0 | 83,337 |
| *PPS+* | Family | 94.0 | 110,664 |
| *PPS* | Family | 92.6 | 97,066 |
| *MEGAN4* | Family | 96.2 | 98,582 |
| *taxator-tk* | Family | 99.8 | 43,751 |
| *PPS+* | Genus | 95.3 | 82,992 |
| *PPS* | Genus | 91.9 | 58,883 |
| *MEGAN4* | Genus | 96.1 | 86,495 |
| *taxator-tk* | Genus | 99.9 | 34,667 |
| *PPS+* | Species | 94.7 | 43,329 |
| *PPS* | Species | N/A | N/A |
| *MEGAN4* | Species | 93.5 | 64,554 |
| *taxator-tk* | Species | 99.7 | 10,314 |

**Figure 10: Comparison of scaffold and contig assignments using *PPS*+ for the chunked cow rumen dataset.**

The comparisons were performed at different taxonomic ranks using heat maps (Sections 4.2.2 and 4.10.1). (panel *a*) Phylum; (panel *b*) class; (panel *c*) order; (panel *d*) family; (panel *e*) genus; (panel *f*) species.

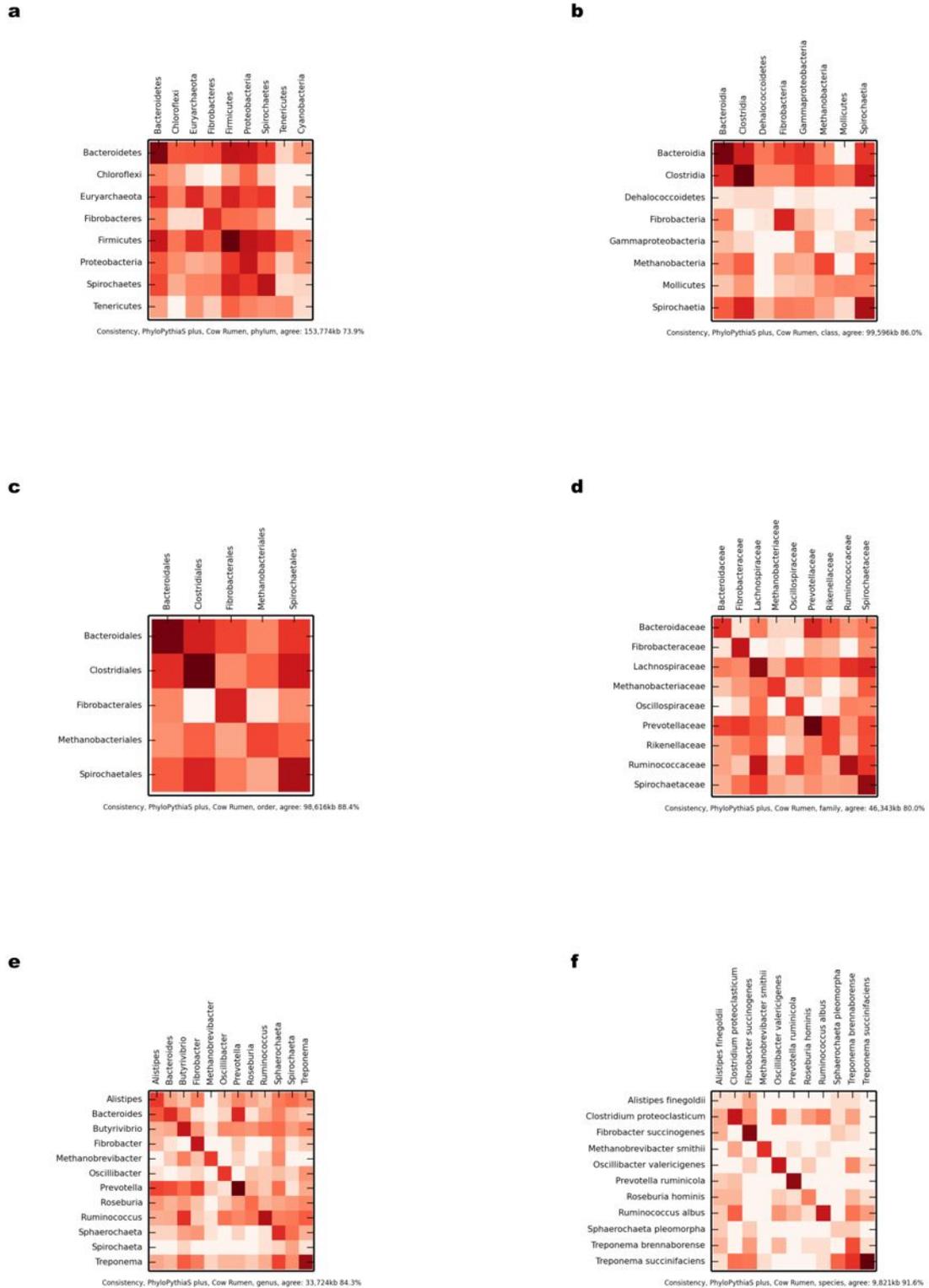

**Figure 11: Comparison of scaffold and contig assignments using the generic *PPS* model for the chunked cow rumen dataset.**

The comparisons were performed at different taxonomic ranks using heat maps (Sections 4.2.2 and 4.10.1). (panel *a*) Phylum; (panel *b*) class; (panel *c*) order; (panel *d*) family; (panel *e*) genus.

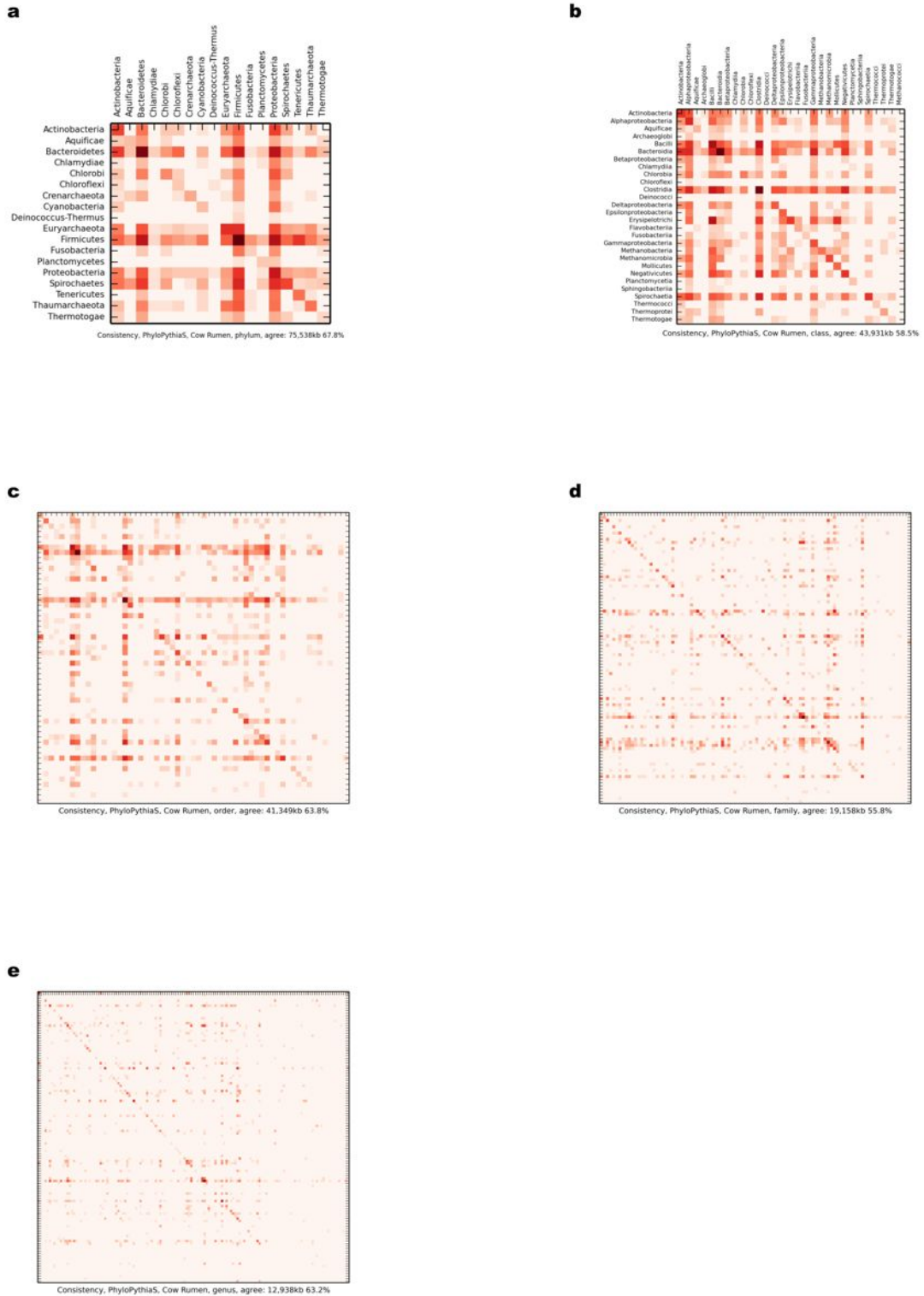

**Figure 12: Comparison of scaffold and contig assignments using *MEGAN4* for the chunked cow rumen dataset.**

The comparisons were performed at different taxonomic ranks using heat maps (Section 4.2.2 and 4.10.1). (panel *a*) Phylum; (panel *b*) class; (panel *c*) order; (panel *d*) family; (panel *e*) genus; (panel *f*) species.

**Figure 13: Comparison of scaffold and contig assignments using *taxator-tk* for the chunked cow rumen dataset.**

The comparisons were performed at different taxonomic ranks using heat maps (Section 4.2.2 and 4.10.1). (panel *a*) Phylum; (panel *b*) class; (panel *c*) order; (panel *d*) family; (panel *e*) genus; (panel *f*) species.

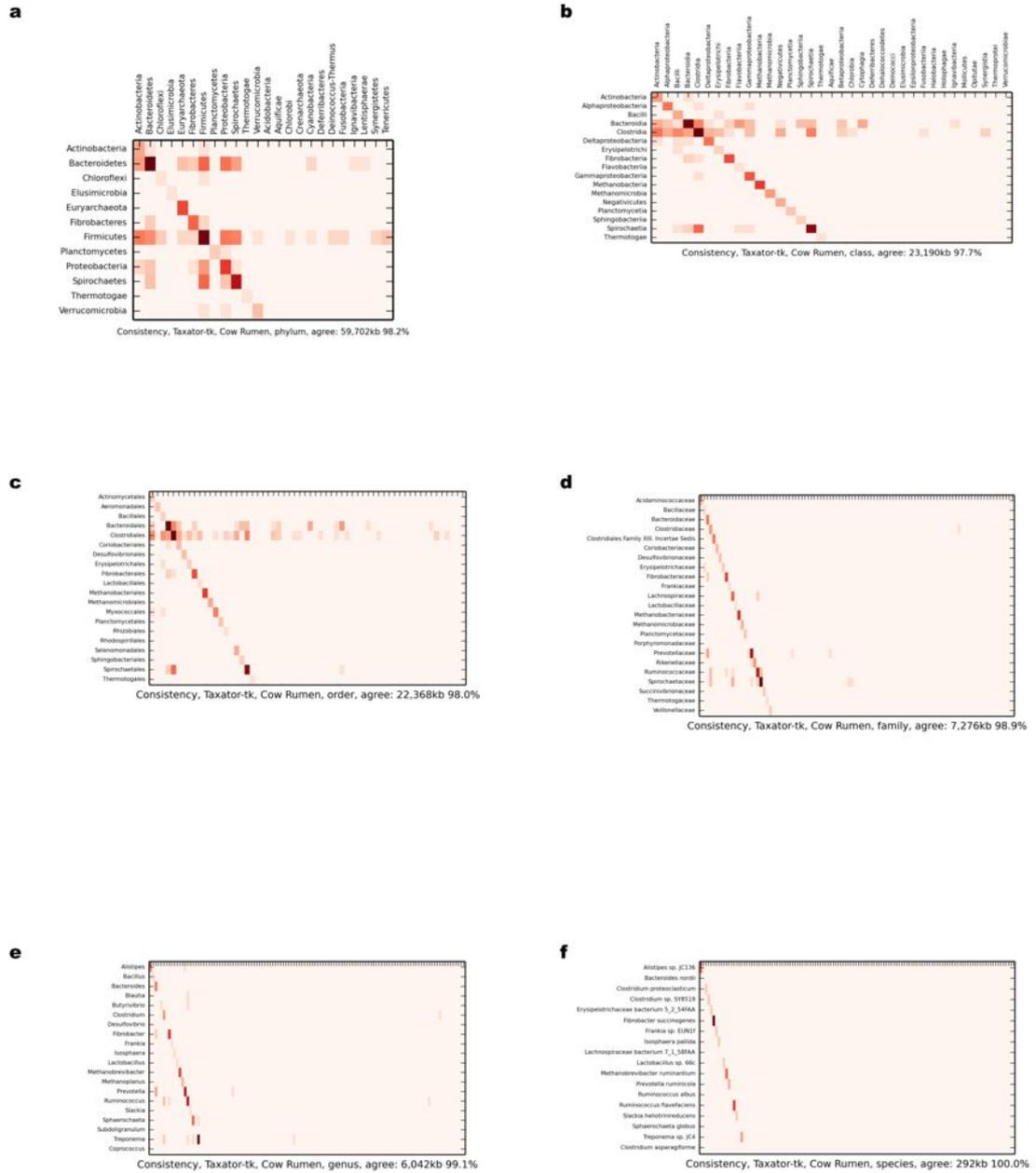

**Figure 14: Comparison of scaffold and contig assignments using *PPS+* for the human gut dataset.**

The comparisons were performed at different taxonomic ranks using heat maps (Sections 4.2.1 and 4.10.1). (panel *a*) Phylum; (panel *b*) class; (panel *c*) order; (panel *d*) family; (panel *e*) genus; (panel *f*) species.

**Figure 15: Comparison of scaffold and contig assignments using the generic *PPS* model for the human gut dataset.**

The comparisons were performed at different taxonomic ranks using heat maps (Sections 4.2.1 and 4.10.1). (panel *a*) Phylum; (panel *b*) class; (panel *c*) order; (panel *d*) family; (panel *e*) genus.

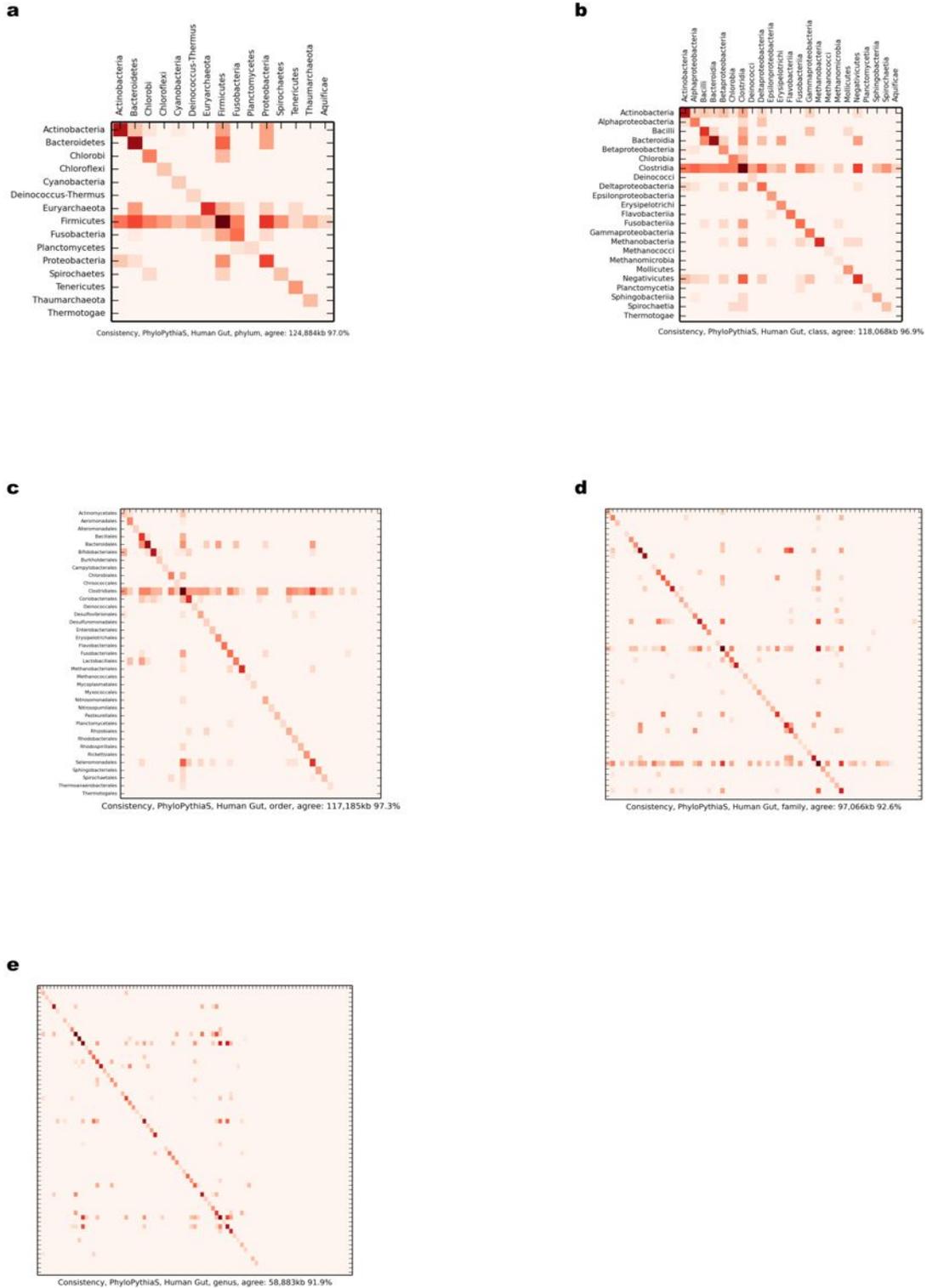

**Figure 16: Comparison of scaffold and contig assignments using *MEGAN4* for the human gut dataset.**

The comparisons were performed at different taxonomic ranks using heat maps (Sections 4.2.1 and 4.10.1). (panel *a*) Phylum; (panel *b*) class; (panel *c*) order; (panel *d*) family; (panel *e*) genus; (panel *f*) species.

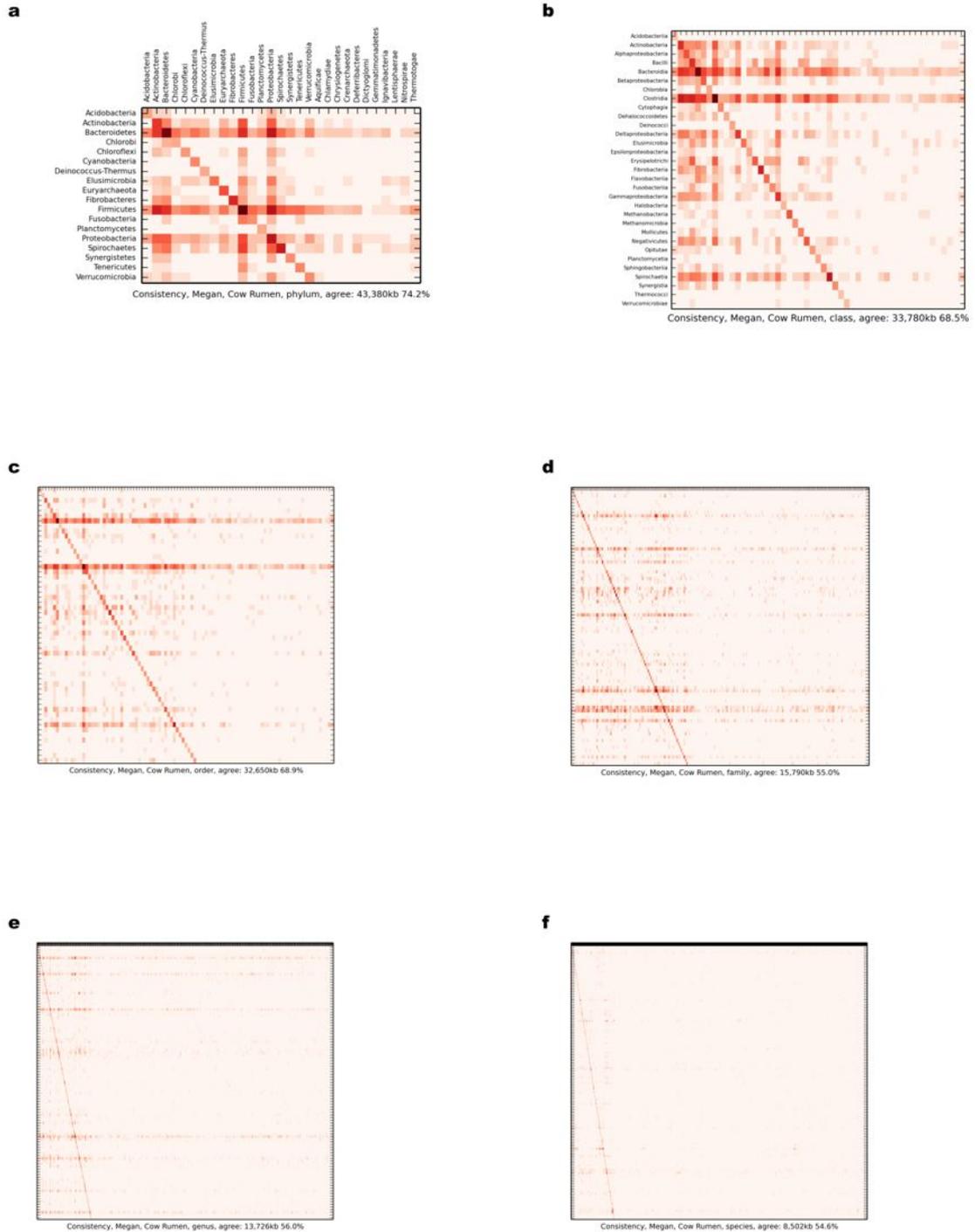

**Figure 17: Comparison of scaffold and contig assignments using *taxator-tk* for the human gut dataset.**

The comparisons were performed at different taxonomic ranks using heat maps (Sections 4.2.1 and 4.10.1). (panel *a*) Phylum; (panel *b*) class; (panel *c*) order; (panel *d*) family; (panel *e*) genus; (panel *f*) species.

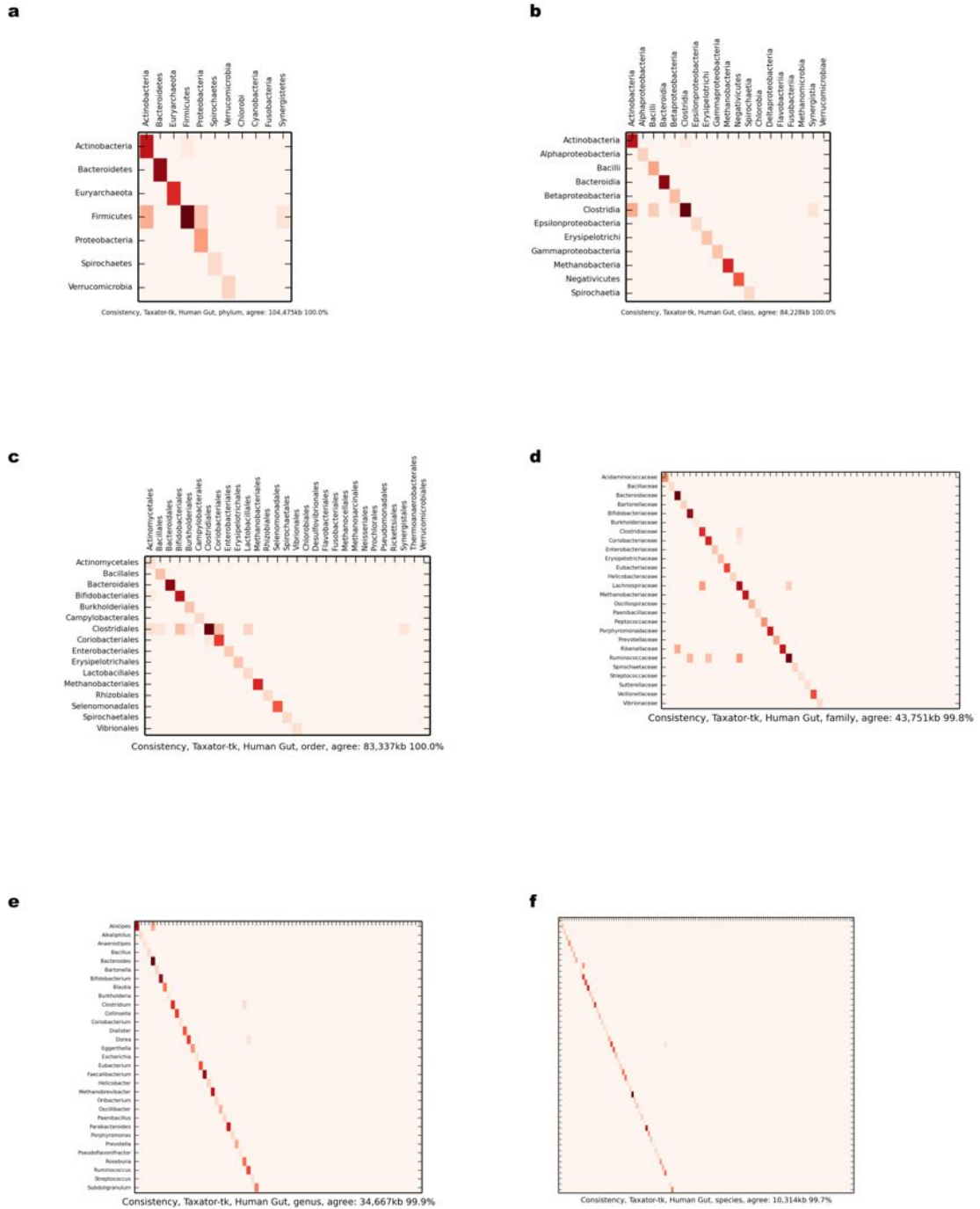

### 5.3.2   Taxonomic scaffold-contig assignment consistency

To assess the quality of taxonomic assignments for these samples, we evaluated the consistency of taxonomic assignments for contigs originating from the same scaffold using a set of measures (Section 4.10.2). These measures assessed the degree to which the taxonomic identifiers of scaffolds and their constituent contigs were consistent relative to each other. This method looked beyond identical identifiers (Section 4.10.1) by taking the relative distances between two taxa in the reference taxonomy into account. The basic idea of these measures is that a scaffold is assigned to a taxonomic identifier of one of its constituent contigs, such that the collective distance of all contig assignments for the respective scaffold to path $p$ in the taxonomy defined by the scaffold identifier is the shortest. The consistency of individual contig assignments is then assessed relative to path $p$: If a contig lies on $p$, it is considered to be assigned consistently; if it does not lie on $p$, it is assigned inconsistently. These measures were computed for the assignments of the chunked cow rumen and the human gut datasets.

Overall, *PPS+* performed better in terms of the consistent assignment of sequences to low taxonomic ranks for the chunked cow rumen dataset and the human gut dataset than the generic *PPS* model and *MEGAN4* (Tables 12 and 13, Def. 6). For both datasets, *taxator-tk* showed the highest consistency according to almost all measures; however, it assigned fewer data to lower taxonomic ranks (family, genus and species) than the other methods.

**Table 12: Scaffold-contig consistency of the chunked cow rumen dataset.**

Scaffold-contig consistency of the assignments made by *PPS+*, the generic *PPS* model, *MEGAN4* and *taxator-tk* for the chunked cow rumen dataset, computed via different definitions (Section 4.10.2). The table also contains the number of kb assigned (kb agreement) at low taxonomic ranks (family, genus and species) and the corresponding consistency (% agreement) (Section 4.10.1). Green numbers correspond to the best values, whereas red numbers indicate the worst values.

| Measure | *PPS+* | *PPS* | *MEGAN4* | *taxator-tk* | Def. |
|---|---|---|---|---|---|
| Scaffolds considered | 12,192 | 12,192 | 9456 | 11,447 | |
| Consistent contigs / total contigs | 128,685 / 159,263 | 137,747 / 159,263 | 116,726 / 135,362 | 151,585 / 153,185 | Def. 1 |
| Consistent count % | 80.80 | 86.49 | 86.23 | 98.96 | Def. 1 |
| Consistent bp / total Kbp | 257,370 / 318,526 | 275,494 / 318,526 | 233,452 / 270,724 | 303,170 / 306,370 | Def. 2 |
| Consistent bp % | 80.80 | 86.49 | 86.23 | 98.96 | Def. 2 |
| Avg. distance to path | 0.38 | 0.30 | 0.50 | 0.02 | Def. 3 |
| Avg. weighted distance to path | 0.38 | 0.30 | 0.50 | 0.02 | Def. 4 |
| Avg. distance to scaffold label | 3.16 | 3.43 | 5.89 | 2.65 | Def. 5 |
| Avg. weighted distance to scaffold label | 3.16 | 3.43 | 5.89 | 2.65 | Def. 6 |
| Family: contigs (kb assigned) | 71,660 | 43,118 | 55,904 | 13,626 | Def. 0a |
| Family: consistency % (scaffolds vs contigs) | 80.0 | 55.8 | 55.0 | 98.9 | Def. 0b |
| Genus: contigs (kb assigned) | 53,705 | 28,077 | 53,008 | 10,596 | Def. 0a |
| Genus: consistency % (scaffolds vs contigs) | 84.3 | 63.2 | 56.0 | 99.1 | Def. 0b |
| Species: contigs (kb assigned) | 26,121 | N/A | 41,204 | 1426 | Def. 0a |
| Species: consistency % (scaffolds vs contigs) | 91.6 | N/A | 54.6 | 100.0 | Def. 0b |

**Table 13: Scaffold-contig consistency of the human gut metagenome dataset.**

Scaffold-contig consistency of the assignments made by *PPS+*, the generic *PPS* model, *MEGAN4* and *taxator-tk* of the human gut dataset computed using different definitions (Section 4.10). Green numbers correspond to the best values, whereas red numbers indicate the worst values.

| Measure | *PPS+* | *PPS* | *MEGAN4* | *taxator-tk* | Def. |
|---|---|---|---|---|---|
| Scaffolds considered | 47,983 | 47,983 | 83,973 | 99,202 | |
| Consistent contigs / total contigs | 64,197 / 66,480 | 63,954 / 66,480 | 99,647 / 101,613 | 117,576 / 117,630 | Def. 1 |
| Consistent count % | 96.57 | 96.20 | 98.07 | 99.95 | Def. 1 |
| Consistent bp / total kbp | 181,207.0 / 189,516.5 | 179,797.6 / 189,516.5 | 191,429.1 / 200,478.2 | 217,517.2 / 217,719.5 | Def. 2 |
| Consistent bp % | 95.62 | 94.87 | 95.49 | 99.91 | Def. 2 |
| Avg. distance to path | 0.06 | 0.07 | 0.05 | 0 | Def. 3 |
| Avg. weighted distance to path | 0.07 | 0.10 | 0.12 | 0 | Def. 4 |
| Avg. distance to scaffold label | 0.63 | 0.72 | 0.38 | 0.29 | Def. 5 |
| Avg. weighted distance to scaffold label | 0.53 | 0.58 | 0.73 | 0.62 | Def. 6 |
| Family: contigs (kb assigned) | 146,046 | 118,679 | 161,452 | 74,793 | Def. 0a |
| Family: consistency % (scaffolds vs contigs) | 94.0 | 92.6 | 96.2 | 99.8 | Def. 0b |
| Genus: contigs (kb assigned) | 110,762 | 71,934 | 149,448 | 61,242 | Def. 0a |
| Genus: consistency % (scaffolds vs contigs) | 95.3 | 91.9 | 96.1 | 99.9 | Def. 0b |
| Species: contigs (kb assigned) | 61,969 | N/A | 114,716 | 20,687 | Def. 0a |
| Species: consistency % (scaffolds vs contigs) | 94.7 | N/A | 93.5 | 99.7 | Def. 0b |

For the chunked cow rumen dataset, the generic *PPS* model assigned more contigs consistently than *PPS+* (Table 12, Def. 2); however this came at the cost of many contigs being assigned to higher taxonomic ranks by *PPS* (Table 12, Defs 0a, 6). *MEGAN4* showed a higher overall consistency than *PPS+* (Table 12, Def. 2) but this was mostly due to many contigs being assigned at higher taxonomic ranks (Table 12, Def. 6). For lower taxonomic ranks or when also taking sequence length into account (instead of the number of assigned sequences), *MEGAN4* was less consistent than *PPS+* (Table 12, Defs 0b, 3–6).

For the human gut dataset, *PPS+* performed better than the generic *PPS* model according to all measures (Table 13, Def. 0–6). *PPS+* was again more consistent than *MEGAN4* when taking sequence lengths into account (Table 13, Defs 2, 4, 6). These measures are more informative for taxonomic binning than the sequence-count based measures (Table 13, Defs 1, 3, 5), as obtaining large bins is desirable. These results also imply that *MEGAN4* assigned substantially more (predominantly short) sequences to lower taxonomic ranks than *PPS+* (Table 13, Def. 0a).

### 5.3.3   Comparison to an expert binning based on marker genes

A taxonomic binning generated by *PhyloPythia* (*PP*) with expert guidance for sample-derived model construction[10] was compared to assignments using the self-training *PPS+*. In this comparison, scaffolds that were unassigned by either method were not considered. The *PP* expert binning and the *PPS+* binning agreed well, down to the order rank (Table 14). For the family and genus ranks, the overlap of both methods dropped to 69.5–74.1%, which may partly be due to parts of the NCBI taxonomy being relabeled since the generation of the expert binning in 2009. The scaffold assignments of both methods were analyzed and compared to assignments based on marker genes (MG) with the + component of *PPS+*. For the MG scaffold assignments, a negligible amount – only two contigs (3.6 kb) of two scaffolds (231 kb) – were used as sample-derived training data for *PPS+*; as mainly sample contigs (2.5 Mb) that were not part of scaffolds were used as sample-derived data to train *PPS*. When comparing *PP* and *PPS+* to the MG assignments, only a small number of scaffolds could have been be compared (7–23% for different ranks), as only some contained marker genes. Both *PPS+* and *PP* assignments were highly consistent with the MG assignments. For instance, at the genus rank, the agreement between *PPS+* and MG was 94.9%, and that between *PP* and MG was 91.6%. Moreover, we compared the number of taxonomic assignments for individual methods (Figure 2). *PPS+* assigned sequences to low-ranking taxa down to the species level, in agreement with the marker gene assignments, while *PP* often assigned the respective sequences to the parental taxa. Only *PP* also included eukaryotes in the model. This demonstrates that *PPS+* can generate high quality taxonomic binning in a fully automated manner.

**Table 14: Comparison to an expert binning based on marker genes.**

Comparison of the taxonomic assignments of *PPS+* versus *PhyloPythia* (*PP*), with expert guidance for sample-derived model construction[10] for the human gut scaffolds (161,343 kb) based on marker genes (*MG*), using the + component of *PPS+*. The measure '% agreement' represents the percentage of bp assigned by both methods to the same taxonomic identifiers at a given rank, whereas 'kb agreement' is the corresponding number of kb assigned by both methods to the same taxonomic identifier. Scaffolds assigned by only one method are not considered in this comparison. Green numbers correspond to the best values, whereas red numbers indicate the worst values.

| Comparison | Rank | % agreement | kb agreement |
|---|---|---|---|
| *PP vs PPS+* | Superkingdom | 99.6 | 160,617 |
| *MG vs PP* | Superkingdom | 99.7 | 38,314 |
| *MG vs PPS+* | Superkingdom | 99.5 | 38,220 |
| *PP vs PPS+* | Phylum | 95.4 | 149,213 |
| *MG vs PP* | Phylum | 96.9 | 17,771 |
| *MG vs PPS+* | Phylum | 98.7 | 18,065 |
| *PP vs PPS+* | Class | 97.0 | 145,887 |
| *MG vs PP* | Class | 98.1 | 17,599 |
| *MG vs PPS+* | Class | 100.0 | 17,869 |
| *PP vs PPS+* | Order | 98.0 | 145,373 |
| *MG vs PP* | Order | 98.3 | 17,494 |
| *MG vs PPS+* | Order | 100.0 | 17,764 |
| *PP vs PPS+* | Family | 69.5 | 95,779 |
| *MG vs PP* | Family | 90.7 | 13,047 |
| *MG vs PPS+* | Family | 83.7 | 12,013 |
| *PP vs PPS+* | Genus | 74.1 | 78,686 |
| *MG vs PP* | Genus | 91.6 | 12,235 |
| *MG vs PPS+* | Genus | 94.9 | 11,479 |

### 5.3.4   Real datasets: evaluation summary

Our evaluation showed that *PPS+* performed substantially better than the generic *PPS* model (Tables 10–13). Moreover, the results of *PPS+* were comparable to a sample-derived model generated according to expert specifications (Table 14). *Taxator-tk* had the highest consistency of all the methods; however, it assigned substantially fewer sequences to low taxonomic ranks than the other methods (Tables 10–13). Our benchmark experiments also confirmed that if the metagenome sequences were closely related to the reference sequences, such as for the human gut dataset, the homology-based methods assigned more sequences correctly to low taxonomic ranks than they did across larger taxonomic distances, as was the case for the cow rumen dataset (Tables 10–13). *PPS+* was not that sensitive to this distance. For *PPS+*, only few taxonomically informative marker genes have to be identified from the sample, for which a substantially larger marker gene reference collection exists than that for genome and draft genome sequences, in terms of the number of species represented in the reference collection. *PPS+* often made more consistent assignments than *MEGAN4* and often assigned the most sequences of all the tested classifiers to lower taxonomic ranks (Tables 10–13).

## 5.4   Throughput comparison

The throughput of the individual methods for contig assignments of the human gut sample was calculated as either Mb or the number of sequences assigned per hour with one thread using the same reference sequences (Sections 4.3 and 4.4). *PPS* and *PPS+* directly use sequences in FASTA format as references, while for *MEGAN4* and *taxator-tk BLAST* or *LAST* databases were initially constructed. Database construction took 6 h 55 m and 81 h 29 min on our servers, respectively for *BLAST* and *LAST*, and was not considered in runtime comparisons. As most time in *PPS+* is spent with model construction, assignment can be further accelerated when reusing models to classify multiple metagenome samples. In this setting, where we consider only the prediction phase of *PPS+*, *PPS+* was more than 7 times faster (up to 0.5 Gb/h) than the homology-based methods (Figure 3). As only a relatively small reference sequence database of 16 Gb was used, runtimes of *BLAST* and *LAST* searches in the homology-based tools would proportionally increase when using larger reference collections.

Unlike the homology-based tools, for which similarity searches require the use of more hardware with more CPUs and main memory, *PPS+* can run on a standard laptop computer. *PPS+* on a laptop with an Intel i5 M520 2.4 GHz processor and 4 GB of RAM was ~1.5–4 times slower than it was on the server with an AMD Opteron 6386 SE 2.8 GHz processor and 512 GB of RAM, mainly due to having insufficient RAM on the laptop, which caused extensive use of the swap space.

# 6   Conclusions

*PPS+* is a taxonomic assignment program that produces accurate assignments with a micro-averaged precision of 80% or more even for low-ranking taxa from metagenome samples with fragments of at least 1 kb. *PPS+* is a fully automated extension of *PPS* and determines the most relevant taxa to be modeled and suitable training sequences directly from the input sample. This enables use of our method for researchers without experience in the field and in studies that generate large collections of metagenome samples. The accurate assignment of sequences of more than 1 kb makes *PPS+* ideally suited for the analysis of assembled datasets generated with common sequencing technologies or the high quality PacBio consensus reads[33].

*PPS+* was substantially faster than other taxonomic binning methods we evaluated. This is partly due to a novel implementation of the *k*-mer counting algorithm, which accelerated *k*-mer counting 100-fold and made *PPS+* up to three times faster than the original *PPS* release. The most time-consuming step remaining in *PPS+* analysis is the inference of the sample-derived models. To further reduce processing times, taxonomic models could be used to classify multiple samples from the same or similar environments.

Due to its speed, *PPS+* is ideally suited for the analysis of large metagenome samples that contain a sufficient number of marker genes. This allows us to model genera and species-level bins and thus partial genome and pan-genome recovery (i.e. a set of sequences assigned to the same bin) from metagenome samples. Like its predecessors, *PPS+* requires only 100 kb of sample-derived data to model a bin, while homology-based methods require large related reference genome or draft genome sequence collections for substantial assignments to low-ranking taxa.

In our evaluation, we did not compare *PPS+* to methods with prohibitive runtimes for large datasets, such as *PhymmBL*, *CARMA3* and *SOrt-ITEMS*. Of the applicable methods, *PPS+* often outperformed *MEGAN4* in terms of micro-averaged precision, micro-averaged recall and consistency. *Taxator-tk* performed best of all in terms of micro-averaged precision and consistency, but assigned substantially fewer sequences to low taxonomic ranks. *PPS+* also excelled in determining taxa that were part of the simulated metagenome community in comparison to the other methods, i.e. *PPS+* produced fewer false taxa.

We found that fully automated *PPS+* binning can be as good as expert-guided binning with the original *PhyloPythia* implementation. *PPS+* also showed substantially improved assignment performance compared to the generic *PPS* model. This can be attributed to the improved match of the training data and modeled taxa to the analyzed metagenome sample. The self-training component introduced in *PPS+* thus allows accurate and fully automated metagenome analysis without manual interventions.

The *PPS+* taxonomic assignment software is available in a virtual machine with a large reference genome sequence collection. This allows metagenome sample analysis on a standard laptop with all common operating systems. It provides end-users with high quality taxonomic binning results for metagenome samples without requiring expensive hardware, manual intervention and expert knowledge.

# *7* References

1.  Pope, P. B. *et al.* Isolation of Succinivibrionaceae implicated in low methane emissions from Tammar wallabies. *Science* **333,** 646–648 (2011).

2.  Patil, K. R. *et al.* Taxonomic metagenome sequence assignment with structured output models. *Nat Methods* **8,** 191–192 (2011).

3.  Imelfort, M. *et al.* GroopM: An automated tool for the recovery of population genomes from related metagenomes. *PeerJ* (2014). doi:10.7287/peerj.preprints.409v1

4.  Albertsen, M. *et al.* Genome sequences of rare, uncultured bacteria obtained by differential coverage binning of multiple metagenomes. *Nat Biotechnol* **31,** 533–538 (2013).

5.  Iverson, V. *et al.* Untangling Genomes from Metagenomes: Revealing an Uncultured Class of Marine Euryarchaeota. *Science* **335,** 587–590 (2012).

6.  McHardy, A. C., Martín, H. G., Tsirigos, A., Hugenholtz, P. & Rigoutsos, I. Accurate phylogenetic classification of variable-length DNA fragments. *Nat Methods* **4,** 63–72 (2007).

7.  Wu, M. & Scott, A. J. Phylogenomic analysis of bacterial and archaeal sequences with AMPHORA2. *Bioinformatics* **28,** 1033–1034 (2012).

8.  Huson, D. H., Mitra, S., Ruscheweyh, H. J., Weber, N. & Schuster, S. C. Integrative analysis of environmental sequences using MEGAN4. *Genome Res* **21,** 1552–1560 (2011).

9.  Dröge, J., Gregor, I. & McHardy, A. C. Taxator-tk: Fast and Reliable Taxonomic Assignment of Metagenomes by Approximating Evolutionary Neighborhoods. *arXiv* 1–18 (2014).

10. Turnbaugh, P. J. *et al.* Organismal, genetic, and transcriptional variation in the deeply sequenced gut microbiomes of identical twins. *Proc Natl Acad Sci USA* **107,** 7503–7508 (2010).

11. Kunin, V., Copeland, A., Lapidus, A., Mavromatis, K. & Hugenholtz, P. A bioinformatician's guide to metagenomics. *Microbiol Mol Biol Rev* **72,** 557–78 (2008).

12. Riesenfeld, C. S., Schloss, P. D. & Handelsman, J. Metagenomics: genomic analysis of microbial communities. *Annu Rev Genet* **38,** 525–552 (2004).

13. Hugenholtz, P. Exploring prokaryotic diversity in the genomic era. *Genome Biol* **3,** REVIEWS0003 (2002).

14. Kalyuzhnaya, M. G. *et al.* High-resolution metagenomics targets specific functional

types in complex microbial communities. *Nat Biotechnol* **26,** 1029–1034 (2008).

15. Hess, M. *et al.* Metagenomic discovery of biomass-degrading genes and genomes from cow rumen. *Science* **331,** 463–467 (2011).

16. Schloissnig, S. *et al.* Genomic variation landscape of the human gut microbiome. *Nature* **493,** 45–50 (2013).

17. Blaser, M. M., Bork, P. P., Fraser, C. C., Knight, R. R. & Wang, J. J. The microbiome explored: recent insights and future challenges. *Nat Rev Microbiol* **11,** 213–217 (2013).

18. Zarowiecki, M. Metagenomics with guts. *Nat Rev Microbiol* **10,** 674–674 (2012).

19. Loman, N. J. *et al.* High-throughput bacterial genome sequencing: an embarrassment of choice, a world of opportunity. *Nat Rev Microbiol* **10,** 599–606 (2012).

20. Metzker, M. L. Sequencing technologies — the next generation. *Nat Rev Genet* **11,** 31–46 (2010).

21. Wu, M. & Eisen, J. A. A simple, fast, and accurate method of phylogenomic inference. *Genome Biol* **9,** R151–R151 (2008).

22. Stark, M., Berger, S. A., Stamatakis, A. & Mering, von, C. MLTreeMap - accurate Maximum Likelihood placement of environmental DNA sequences into taxonomic and functional reference phylogenies. *BMC Genomics* **11,** 461–461 (2009).

23. Sunagawa, S. *et al.* Metagenomic species profiling using universal phylogenetic marker genes. *Nat Methods* **10,** 1196–1199 (2013).

24. Segata, N. N. *et al.* Metagenomic microbial community profiling using unique clade-specific marker genes. *Nat Methods* **9,** 811–814 (2012).

25. Dröge, J. & McHardy, A. C. Taxonomic binning of metagenome samples generated by next-generation sequencing technologies. *Brief Bioinform* (2012).

26. Gerlach, W. & Stoye, J. Taxonomic classification of metagenomic shotgun sequences with CARMA3. *Nucleic Acids Res* **39,** e91–e91 (2011).

27. Monzoorul Haque, M., Ghosh, T. S., Komanduri, D. & Mande, S. S. SOrt-ITEMS: Sequence orthology based approach for improved taxonomic estimation of metagenomic sequences. *Bioinformatics* (2009).

28. Methé, B. A. *et al.* A framework for human microbiome research. *Nature* **486,** 215–221 (2012).

29. Karlin, S. & Burge, C. Dinucleotide relative abundance extremes: a genomic signature. *Trends Genet* **11,** 283–290 (1995).

30. Deschavanne, P. J., Giron, A., Vilain, J., Fagot, G. & Fertil, B. Genomic signature:

characterization and classification of species assessed by chaos game representation of sequences. *Mol Biol Evol* (1999).

31. Brady, A. & Salzberg, S. PhymmBL expanded: confidence scores, custom databases, parallelization and more. *Nat Methods* **8,** 367–367 (2011).

32. Rosen, G. L., Reichenberger, E. R. & Rosenfeld, A. M. NBC: the Naive Bayes Classification tool webserver for taxonomic classification of metagenomic reads. *Bioinformatics* **27,** 127–129 (2011).

33. Chin, C. S. *et al.* Nonhybrid, finished microbial genome assemblies from long-read SMRT sequencing data. *Nat Methods* **10,** 563–569 (2013).

34. Carr, R., Shen-Orr, S. S. & Borenstein, E. Reconstructing the Genomic Content of Microbiome Taxa through Shotgun Metagenomic Deconvolution. *PLoS Comput Biol* **9,** e1003292–e1003292 (2013).

35. Joachims, T., Finley, T. & Yu, C. N. Cutting-plane training of structural SVMs. *Mach Learn* **77,** 27–59 (2009).

36. Pope, P. B. P. *et al.* Metagenomics of the svalbard reindeer rumen microbiome reveals abundance of polysaccharide utilization Loci. *PLoS ONE* **7,** e38571–e38571 (2011).

37. Eddy, S. R. Accelerated Profile HMM Searches. *PLoS Comput Biol* **7,** e1002195–e1002195 (2011).

38. Cole, J. R. *et al.* The Ribosomal Database Project: improved alignments and new tools for rRNA analysis. *Nucleic Acids Res* **37,** D141–D145 (2009).

39. Schloss, P. D. *et al.* Introducing mothur: open-source, platform-independent, community-supported software for describing and comparing microbial communities. *Appl Environ Microb* **75,** 7537–7541 (2009).

40. Huang, Y., Gilna, P. & Li, W. Identification of ribosomal RNA genes in metagenomic fragments. *Bioinformatics* **25,** 1338–1340 (2009).

41. Karp, R. M. & Rabin, M. O. Efficient randomized pattern-matching algorithms. *IBM J Res Dev* **31,** 249–260 (1987).

42. Marcais. A fast, lock-free approach for efficient parallel counting of occurrences of k-mers. *Bioinformatics* 764–770 (2011).

43. Hu, X. *et al.* pIRS: Profile-based Illumina pair-end reads simulator. *Bioinformatics* **28,** 1533–1535 (2012).

44. Huang, W., Li, L., Myers, J. R. & Marth, G. T. ART: a next-generation sequencing read simulator. *Bioinformatics* **28,** 593–594 (2012).

45. Richter, D. C. D., Ott, F. F., Auch, A. F. A., Schmid, R. R. & Huson, D. H. D.

MetaSim: a sequencing simulator for genomics and metagenomics. *CORD Conference Proceedings* **3,** e3373–e3373 (2007).

46.    Mavromatis, K. *et al.* Use of simulated data sets to evaluate the fidelity of metagenomic processing methods. *Nat Methods* **4,** 495–500 (2007).

47.    Debruijn, I. MetAssemble. *github.com* (2014). at <https://github.com/inodb/metassemble/>

48.    Zerbino, D. R. D. & Birney, E. E. Velvet: algorithms for de novo short read assembly using de Bruijn graphs. *Genome Res* **18,** 821–829 (2008).

49.    Treangen, T. J. T., Sommer, D. D. D., Angly, F. E. F., Koren, S. S. & Pop, M. M. Next generation sequence assembly with AMOS. *Curr Protoc Bioinformatics* **Chapter 11,** 11.8.1–11.8.18 (2011).

50.    Luo, R. *et al.* SOAPdenovo2: an empirically improved memory-efficient short-read de novo assembler. *GigaScience* **1,** 1–6 (2012).

51.    Namiki, T., Hachiya, T., Tanaka, H. & Sakakibara, Y. MetaVelvet: an extension of Velvet assembler to de novo metagenome assembly from short sequence reads. *Nucleic Acids Res* 1–12 (2012).

52.    Roche. Newbler. (2014). at <http://www.454.com/products/analysis-software/>

53.    Camacho, C. *et al.* BLAST+: architecture and applications. *BMC Bioinformatics* **10,** 421–421 (2009).

54.    Federhen, S. S. The NCBI Taxonomy database. *Nucleic Acids Res* **40,** D136–D143 (2011).

55.    Sayers, E. W. *et al.* Database resources of the National Center for Biotechnology Information. *Nucleic Acids Res* **37,** D5–15 (2008).

56.    Pruesse, E. *et al.* SILVA: a comprehensive online resource for quality checked and aligned ribosomal RNA sequence data compatible with ARB. *Nucleic Acids Res* **35,** 7188–7196 (2007).

57.    Maglott, D. D., Ostell, J. J., Pruitt, K. D. K. & Tatusova, T. T. Entrez Gene: gene-centered information at NCBI. *Nucleic Acids Res* **33,** D54–D58 (2004).

58.    Frith, M. C., Hamada, M. & Horton, P. Parameters for accurate genome alignment. *BMC Bioinformatics* **11,** 80 (2010).